\documentclass[12pt,preprint]{aastex}

\shorttitle{Contact Binaries with Additional Components.I}
\shortauthors{Pribulla \& Rucinski}

\begin{document}

\title{Contact Binaries with Additional Components.~I\\
The Extant Data}

\author{Theodor Pribulla}
\affil{Astronomical Institute of the Slovak Academy of Sciences\\
059~60 Tatransk\'a Lomnica, Slovak Republic}
\email{pribulla@ta3.sk}

\and

\author{Slavek M. Rucinski}
\affil{David Dunlap Observatory, University of Toronto \\
P.O.~Box 360, Richmond Hill, Ontario, Canada L4C~4Y6}
\email{rucinski@astro.utoronto.ca}

\begin{abstract}
We have attempted to establish an observational evidence for
presence of distant companions which may have acquired and/or
absorbed the angular momentum during evolution of multiple systems
thus facilitating or enabling formation of contact binaries.
In this preliminary investigation we use several 
techniques (some of them distance-independent) and
mostly disregard detection biases of individual techniques
in an attempt to establish a lower limit to the frequency
of triple systems. While the whole sample of 151 contact
binary stars brighter than $V_{max} = 10$ mag.\ gives a firm
lower limit of 42\% $\pm$ 5\%, the corresponding number
for the much better observed Northern-sky sub-sample
is 59\% $\pm $8\%. These estimates indicate that most
contact binary stars exist in multiple systems.
\end{abstract}


\section{INTRODUCTION}
\label{intro}

Formation of binary stars is a fascinating subject generating
a very active research effort. Separate conferences \citep{zinn2001}
are devoted to this subject while specialized review
papers \citep{tholine2002,zinn2002,bate2004}
describe difficulties in understanding of this complex
process. Binaries with orbital periods of the order of
a hundred of days are usually considered as close in this
context. We have no idea how short-period
binaries with periods much shorter than
3 to 5 days form. In fact, such binaries, particularly
those with periods shorter than one day, should not exist:
Indeed, even if some unknown process would form contact
binaries at the T~Tauri stage, relatively large sizes
of component stars would imply the resulting orbital
periods to be longer than about 3 -- 5 days.

Formation in triple (or multiple) systems may
alleviate the close-binary formation difficulty by allocating most of
the angular momentum in the most distant component of a
triple/multiple system, leaving a low angular momentum remnant.
Such a scenario would have important
observational consequences because -- by observing close binaries --
we would preferentially select stars with
distant companions. One particular mechanism involving
the Kozai cycles \citep{koza1962} may produce close binaries by
invoking a very strong tidal interaction during periastron passages
when the inner, close (low angular momentum) orbit is perturbed by the
third body into a very eccentric one. The close periastron
interactions may then lead to a rapid angular momentum loss
\citep{kisel1998,EggKis2001}.

A confirmation of the triple body formation of very close binaries
can only come through observations and careful statistical investigations.
Unfortunately, proving or disapproving this hypothesis is difficult
and -- as for all statistical studies -- is hindered by
observational bias and selection effects. In this paper,
we limit our scope to contact binaries,
the most extreme objects with the
lowest angular momentum content among all Main Sequence (MS) binary stars.
Contact binaries \citep{Rci1993}
are made of solar-type stars with spectral types
ranging from early K, through G and F to middle-A; they have
short orbital periods ranging between 0.22 day
to about 0.7 day at which
point the statistics becomes unreliable and properties
of massive contact binaries less well defined. Thus, in spite
of the unquestioned existence of very massive contact binaries,
it is not clear if a continuous sequence joins the solar-type
contact binaries (stars also known as W~UMa type variables)
with the early-type B and O-type binaries. In this paper,
we limit ourselves to binaries with periods shorter than
1 day which is the traditional period limit for the solar-type
(W~UMa-type) contact binaries.

\citet{RK1982} commented on a high incidence of visual binaries
among contact binaries while discussing a visual binary
consisting of two W~UMa-type systems, BV~Dra and BW~Dra.
Later on, \citet{Chambliss1992}
listed several cases of triple systems with contact binaries
in his compilation and \citet{HenryMki1998} detected a few cases of
triple systems. However, these
were fragmentary, almost anecdotal observations and there clearly exists
a need of quantifying the matter of the multiplicity of very close binaries.

This paper attempts to characterize all indications of triplicity and/or
multiplicity by categorizing contact systems into those with
``detected'', ``suspected'' and ``non-detected'' companions,
with an additional obvious category of ``not observed'' ones.
The techniques actually used are described in Sections~\ref{direct}
-- \ref{indicators}.
Not all techniques discussed by us are equally useful and not
all imaginable techniques have been considered; there is still
a lot of room for expansion. Also, types of companions
are not explicitly discussed in the paper as our goal was
simply to summarize the extant information irrespectively of what
accompanies a given contact binary. Thus, a huge subject
of selection effects and biases is only glanced over in this paper.
Future studies, such as Paper~II of this series on spectroscopic
detection in spectra used for the DDO radial velocity program
\citep{dangelo2005}, will more fully address
specific limitations of various detection techniques.

We describe our sample in Section~\ref{sample}.
The main results of this paper are described in
Sections~\ref{direct} -- \ref{indicators} and
summarized in Table~\ref{summary} where we collect all partial results
and attempt to combine ``suspected'' cases into detections.
Results for individual cases are summarized in a condensed form
in the column FLAGS by codes specific for each technique.
The codes are explained in Table~\ref{flags} while
individual binaries requiring comments are listed in
Table~\ref{notes}. Section~\ref{conclude} summarizes the
main results of this study while Section~\ref{future} gives
a discussion of possible directions for the future work.

\section{THE SAMPLE}
\label{sample}

Meaningful statistical results on stellar properties, including
the matter of multiplicity, can only be properly studied utilizing
volume limited samples. Unfortunately, a proper volume-limited
sample of contact binaries does not exists. In fact, as
we describe below, we are very far from having
even a good magnitude-limited sample.

As usually for stellar statistics based on diversified data,
strengths and biases of various techniques tend to depend on
brightness. We decided to set a limit for
our sample at the magnitude $V_{max}=10$. This
magnitude limit is a compromise between a sufficient size of the
sample and a reasonable completeness. At this point we feel
confident that the only complete sample of contact binaries with
variability amplitudes $>0.05$ mag.\ is that of the Hipparcos
Catalogue \citep{Rci2002b} which extends to $V_{max}=7.5$. From
now on, we will call it ``the Hipparcos sample'' or ``the 7.5
mag.\ sample''. Beyond that limit, the Hipparcos target selection
was driven by lists existing in the early 1980's, so that the
Hipparcos catalog is incomplete.

The original 7.5 mag.\ sample, as defined in
\citet{Rci2002b},  consists of 32 stars, in
this 13 genuine contact binaries (EW), 14
ellipsoidal variables of small amplitude (EL) and 5 binaries with
unequally deep minima (EB).
Since small amplitude contact binaries are
expected to dominate in numbers \citep{Rci2001}, it is
reasonable to consider the EL and the EW systems
together. It is not obvious if we should add to these the
EB systems \citep{Rci2002b}, but their number is relatively
small. For the purpose of this paper, we decided to 
include in the 7.5 mag.\ sample three triple
systems (V867~Ara, KR~Com and HT~Vir), where the contact binary
is slightly, by less than 0.1 mag.,
fainter than $V_{max}=7.5$, but the total brightness
beyond that limit led previously to their elimination from the
sample. This is consistent with the ``fuzzy'' $V_{max}=10$ limit
that we adopt in this paper, as described below.

Assuming a simple scaling law for the uniform, homogeneous
stellar density, we can expect that the 10 mag.\ sample should
be $\simeq 32 \times$ larger than the 7.5 mag.\ sample.
Thus, it should have about one thousand binaries. In fact, the 10 mag.\
sample consists of 151 objects which directly shows that huge
numbers of contact binaries still remain to be discovered in the
interval $7.5 < V_{max} < 10$. A part of this discrepancy
is probably the detection threshold for variability
amplitudes which is about 0.05 mag.\ for the Hipparcos sample,
but is not uniform and possibly at a level of 0.1 -- 0.2 mag.\
for the 10 mag.\ sample.

The number of 35 binaries in the Hipparcos sample is small for
meaningful statistical inferences, but it can provide some
guidance on the triple-system incidence for the $V_{max}=10$ sample.
Utilizing techniques identical to those described in detail
in this paper, we see 17 triple systems which
corresponds to the lower limit on the
relative incidence of triple systems of $48 \pm 12$\%. As we will
see in Sections~\ref{conclude} -- \ref{future}, this
is consistent with, but slightly lower than 
the results for the full $V_{max}$ = 10 mag.\ sample indicating 
this fraction to be  59\% $\pm$ 8\%. We note that a large
fraction of the Hipparcos sample (about 1/3) are actually low
photometric amplitude binaries discovered by this satellite; these
binaries have been studied only during the last 7 -- 8 years and
still have incomplete data.

We note that we encountered several difficulties
in observing the strict limit of 10 mag.
In particular, our definition of $V_{max}$ refers to
the combined magnitude of all components in the system,
irrespectively if this is just a binary or has a companion.
Since we a priori do not know how many stars are in the system,
such a ``fuzzy'' definition explicitly
depends on the number of components in
the system. While we recognize this uncertainty, we note that
a comparable uncertainty of a few 0.1 mag.\
in the magnitude limit is currently introduced by
differences between photometric systems
and a generally poor state of uniform,
calibrated photometric data for contact
binaries. In the practical implementation of the
$V_{max}=10$ limit we have been forced to use several sources.
We usually took $V_{max}$ from the General Catalogue of
Variable stars (hereafter GCVS)\footnote{We used the most recent
electronic version 4.2 available
at http://www.sai.msu.su/groups/cluster/gcvs/}.
For stars without $V_{max}$ magnitudes in the GCVS, we used
5th percentile of the Hipparcos brightness (H49 field in the
Hipparcos Catalog \citep{hip}); these $H_p$ magnitudes
were then transformed to Johnson photometric systems by interpolating
in the tables of $H_p - V_J$ (Tables 1.3.5. and 1.3.6 in Volume~1 of the
Hipparcos Catalogue). For a few stars not observed during
the Hipparcos mission and having only photographic or photo-visual
maximum magnitude in GCVS, we took approximate $V_{max}$ from
original papers.

Finally, we want to note that we resisted very hard a temptation
to use other samples in addition to the 10 mag.\ sample. Some
sources provided deeper samples. For example, a sample based on
the Hipparcos astrometric data would consist of 177 systems and --
for some binaries -- would go well beyond the adopted magnitude
limit. On the other hand, the
Hipparcos sample is incomplete before the 10 mag.\ limit is
reached and some bright systems are not included or
have poor data. Other techniques and methods that we used had different
biases and limitations so that we would have to analyze each
sample separately. In the end, we decided to strictly observe the limit of
$V_{max} = 10$ which results in the final sample of 151 binaries.
We list binaries fainter than 10 mag.\ separately in the second
half of Table~\ref{summary}, but we do not utilize them for the
multiplicity frequency estimates nor do we show them in frequency
plots. They are included here partly for the record, with an
expectation that the magnitude limit will soon be improved, partly
to show that different techniques and multiplicity indicators have
different properties at faint brightness levels.
We note that Table~\ref{summary} may be
taken as a useful list for further work on candidates of
multiplicity.

\section{DIRECT IMAGING AND ASTROMETRY}
\label{direct}

\subsection{The WDS Catalog and Hipparcos data (FLAG-1)}
\label{WDS}

The most obvious of all detection techniques is the direct
visual detection utilizing speckle interferometry or
high-resolution imaging (see e.g., \citet{Hershey1975}).
A large body of data for multiple stars has been collected by
\citet{Tok1997,Tok2004}; his comprehensive Catalogue of Physical
Multiple Stars provides data on such quantities as the
brightness, distance, position angle, astrometric data and
hierarchy in multiple systems. These data have
been used in the present study together with the
Hipparcos Catalogue \citep{hip} data, augmented by the
post-Hipparcos improved orbits of many wide systems by \citet{Sod1999},
with component separation and photometric data
revised by \citet{Fabmak2000}.

A visual detection should involve confirmation
of the physical link, which is usually relatively simple for
contact binaries as -- photometrically -- they occur within the width of the
Main Sequence so that a plausible combination
of magnitude and color differences is usually sufficient to establish
a physical connection. The main problem here is the
lack of standardized photometric data for a large fraction
of contact systems.
Bright -- hence nearby -- contact binaries appear randomly
on the sky, usually on an empty sky so that the presence of
a close visual companion can be frequently taken as
an indication of the triplicity, obviously requiring confirmation.

Visual detections and astrometric data have been
accumulated over a long time as a result of many
investigations using different instruments and methods.
The Washington Double Star (WDS) Catalog \citep{WDS} appears to
be the most complete and best updated compilation
of these results; the major contributor to the WDS is the
Hipparcos satellite mission \citep{hip}. One of great contributions
of this satellite, in addition to the well-known
astrometric results, was in the photometric discovery of many
low amplitude, $0.05 < \Delta V < 0.2$ mag., variables. Inclusion
of these variables dramatically improved the available data on
the statistics of close binaries \citep{Rci2002b}
after appropriate allowance for binaries misclassified as pulsating
variables, particularly after a major effort to identify those
unrecognized contact binaries by \citet{duerb1997}.

Among the 137 systems listed in the Hipparcos Catalogue, 43 are
listed in WDS to form visual binaries with a distant third body,
although some (e.g., AW~UMa, see Section~\ref{conclude})
are in proper-motion pairs with very long, practically
un-measurable orbital periods, yet at physical separations
indicating a common origin. Multiplicity of a few systems
(V867~Ara, QW~Gem, CN~Hyi, V1387~Ori, Y~Sex, and TU~UMi) was
discovered by the Hipparcos mission while multiplicity of CV~Cyg
and EF~Dra was discovered by the TYCHO project of that mission
(see The Tycho Double Star Catalogue: \citet{fabr2002}).

In order to avoid a distance bias, instead of the
angular sky separation ($\rho$) as a criterion of a physical
link, we utilized distances
to find the projected separations in AU. For systems having
Hipparcos parallaxes ($\pi$) with relative error less than 20\%,
the separations were determined as $\rho/\pi$. Distances to systems
without reliable parallaxes were determined using the
absolute magnitude calibration of \citet{rd1997} which is
based on the orbital period and color index; for a
few systems, the color index had to be estimated from the
spectral type or even more crudely from the orbital period.
For EM~Cep and V395~And, the $(B-V)$ index is too blue for the
calibration to apply. For EM~Cep a distance of $d$ = 794 pc  was
adopted assuming its membership in NGC7160 \citep{khar2005}.
The spectral type of B7--8 for V395~And with the
Main Sequence components implies a distance of 450pc, but
this is an upper limit as the components can be
sdB stars \citep{ddo10}.

The direct visual detections documented in the WDS are marked in
Table~\ref{summary} by the first of the eight flags (FLAG-1),
which can take the following designations:
``D'' for detected (separations less than 2,000 AU or 0.01 pc),
``P'' for possible cases (separations between 2,000 and
20,000 AU corresponding to 0.01 to 0.1 pc), and
``N'' for separations larger than 20,000 AU.

% Table 1
\placetable{summary}

\subsection{Adaptive optics CFHT program (FLAG-2)}
\label{sec_cfht}

A program of direct detection of infrared companions using
the adaptive optics system on the Canada France Hawaii Telescope
(CFHT) was undertaken by one of the authors (SMR) on two
nights in 1998. It consisted of one-color, one-epoch, high spatial
resolution imaging to $<0.1$ arcsec.
The data have been analyzed in a cursory way because
continuation of the program is expected so that
the results should be considered as preliminary.

The observations were obtained with the PUEO instrument and the
KIR camera combination
\citep{rigaut1998} in the $H$ and $K$ infrared bands (1.65 and 2.2 $\mu m$)
on January 10/11 and July 23/24, 1998. Most observations were done in
the $K$ band and its narrow band version, the $K_{CO}$ band,
with a few binaries observed also in the $H$ band. Fields of $36
\times 36$ arcsec were observed, centered on contact binaries
visible from CFHT to approximately $V < 10.5$. The scale on the
detector was 0.0348 arcsec/pixel. The measured FWHM of corrected images
was 0.143 arcsec in the $K$ band, which is very close to the expected
diffraction limited performance of a 3.6 meter telescope. We used the PSF
subtraction only for obvious, easily visible
companions at separations below 1~arcsec;
we did not apply this technique to search for new, faint companions perhaps
hiding in the diffraction ring structure of the primary -- this may be
attempted in the future. In spite of the
cursory treatment, we detected seven previously not identified
companions at separations smaller than 5 arcsec to
GZ~And, AH~Aur, CK~Boo, SW~Lac, V508~Oph, U~Peg and RZ~Tau.
In all those cases, judging by the values of magnitude
differences $\Delta K$ and $\Delta H$,
we suspect the companions to be late-type dwarfs.
We estimate that our detection limit in the sense of the smallest
separations is about 0.08 arcsec (for a system
with identical components); this is based on the case of CK~Boo
where a fainter companion was easily to detect at
a separation of 0.12 arcsec.

\placetable{cfht}
% Table 2: CFHT results

Some of the observed fields contain close companions at moderate
angular separations ($5 \le \rho \le 25$ arcsec) which may be
physically associated with the central contact binaries because
such angular separations for nearby stars still imply physical
distances between stars well below the typical average in the solar
neighborhood. These
cases must be supported by other evidence which we currently do
not have. Results of the CFHT observations are listed in
Table~\ref{cfht}. We give the separation in arcsec, the position
angle in degrees and the difference in magnitudes relative to the
primary (contact binary) component in the $K$ or $H$ bands.
The final assessment of the physical connection is based
-- similarly as for the directly observable visual binaries
(Section~\ref{WDS}) --
on the projected separation of the components. In
Table~\ref{summary}, the results are coded by FLAG-2, similarly
as for objects discussed in the previous Section~\ref{WDS},
except for the symbol ``--'' which signifies
that the star was not observed.

\subsection{Detection through astrometric solutions (FLAG-3)}

The most important astrometric results and the reason for
the Hipparcos mission were trigonometric parallaxes of nearby stars.
Parallax errors and quality estimates of astrometric and orbital
solutions are cataloged in the Hipparcos Double and Multiple Annex
for all known or discovered visual pairs. We present here an
argument that the parallax error data carry useful information on
multiplicity.

The Hipparcos parallax error depends primarily on the brightness and
less so on the ecliptical latitude of a system, but it
increases if an unseen component causes a transverse motion of the
star on the sky. The astrometric solution can be also deteriorated by a
variability-induced motion of the photocenter in a close visual or
unresolved system with a variable component; such a
motion can be particularly pronounced for large amplitude variables like Mira
pulsating stars (\citet{pourb2003}), but may also affect all other
variables, including contact binaries. In what follows we analyze
the parallax errors as indicators of multiplicity
of contact binaries\footnote{One of notable multiple systems
without a parallax is GZ~And. Together with a few similar stars
close by in the sky it appears to form a genuine
multiple-star trapezium configuration \citep{walk1973}.}.
For an assessment of the multiplicity, the most important Hipparcos
catalog data are:
\begin{itemize}
\item The astrometric solution type (the Catalog field code H59); the
first character in the column HIP in
Table~\ref{summary}\footnote{Two other Hipparcos H59 flags do not
occur within our sample: ``O'' astrometric orbital solution
and ``V'' variability induced motion; also a ``D'' type of visual solution
has never occurred.}. It may take values of ``C'' for
a linear motion of components, ``X'' for a
stochastic solution, ``G'' for an acceleration term being necessary.
While ``X'' and ``G'' apply for apparently single 
stars\footnote{By a ``single'' star in this context, 
we mean a contact binary without 
additional components directly resolved on the sky. 
Visual binaries with one component being a contact binary are 
therefore -- strictly speaking -- triple systems.}
(6 cases in our sample), ``C'' refers to known visual systems.
\item The quality of the double or multiple star solution
(the field code H61). This is indicated by
the second symbol in the column HIP in
Table~\ref{summary}.  The flag ``S'' indicates that a
possible, although unconvincing, non-single
solution was found for an apparently
single star; 17 such systems appear in our sample.
For known visual binaries, the solutions are coded by ``A'' to ``D'',
in progression from best to worst. A very poor quality of the
solution (i.e., the flags ``C'' or ``D'') indicates a possible
additional perturbation in the visual system.
\item The Hipparcos parallax error (field H16); column $\sigma_\pi$
in Table~\ref{summary}. 35 systems have parallax errors more
than 3--$\sigma$ larger than respective median
values at given brightness levels.
\end{itemize}

An inspection of the catalog shows that three
contact systems, V353~Peg, ER~Ori, and V2388~Oph, required an
acceleration term in the astrometric solution, hence they are most probably
astrometric double stars with orbital periods longer than about 10 years
(see Section 2.3.3 of the Hipparcos Catalogue). The
acceleration terms in astrometric solution of
ER~Ori are relatively large:
$g_\alpha = -19.26\pm$6.57 mas~yr$^{-2}$ and
$g_\delta = -17.34\pm$4.71 mas~yr$^{-2}$
(the unit ``mas'' to be used hereinafter is equal
0.001 of arcsec, for a milli-arcsec). Several systems, which are
very probably unresolved visual binaries, show ``stochastic'', i.e.,
an excessive random scatter in deviations from their
astrometric solutions (CL~Cet, BE~Scl, FI~Boo, AK~Her, V401~Cyg,
and CW~Sge\footnote{Note that V401~Cyg, CW~Sge and BE~Scl as well
as UX~Eri and V1363~Ori (the next paragraph) are slightly
fainter than 10 mag and thus appear in the second half of
Table~\ref{summary}}). 
The so-called ``cosmic error'' $\epsilon$ given in the
Hipparcos catalog describes the scatter in the
astrometric solution. It is of considerable astrophysical
interest as it gives the most probable size
of the astrometric orbit and points at relatively tight orbits
with periods shorter than one year. The cosmic error is the largest for
V401~Cyg with $\epsilon = 26.82 \pm 2.24$ mas and BE~Scl where
$\epsilon = 13.36 \pm 1.35$ mas. The systems are prime candidates
for speckle interferometric observations.
In fact, \citet{mason99,mason01} performed speckle interferometry
of Hipparcos doubles and Hipparcos problem stars (i.e.\ stars
having ``V'', ``O'', ``G'', ``X'' in H59 or ``S'' in H61)
and detected 15 new visual doubles with a
mean separation about 0.2 arcsec. Among these stars was
V353~Peg (``G'' flag in H59) but no companion of this star
at an angular separation larger than 0.035 arcsec
was detected \citep{mason01}.

The parallax error vs.\ Hipparcos magnitude  $H_P$ is
plotted in Figure~\ref{errpx-V}. The figure clearly shows that
known visual binaries (marked by filled circles)
show large parallax errors, a condition
which can be used as a criterion indicating multiplicity.
The large parallax errors correspond also to stochastic solutions
or to cases where necessity of acceleration term occurred (symbols
in the figure framed by a square). Of note is the case
of ER~Ori with a formally {\it negative\/} parallax
$\pi = -6.68$ mas. In addition to systems
already indicated in the Hipparcos catalogue,
large parallax errors of DY~Cet, UX~Eri and V1363~Ori strongly
suggest multiplicity.

% Fig.1: parallax error versus magnitude
\placefigure{errpx-V}

A summary of all multiplicity indications derived from the
astrometric parallax data appears in Table~\ref{summary} under FLAG-3.
Detections are marked by ``D'' for ``X'' or ``G'' type solutions, while
the suspected cases ``S'' correspond to those already suspected
in the Hipparcos catalog (``S'' in H61), systems with
``C'' type quality of orbital solutions\footnote{This code may indicate
an additional disturbance in the orbital solution and a higher multiplicity.}
and systems with parallax errors more than 3--$\sigma$ from
the median error (where $\sigma$ is the $rms$ uncertainty in the
median error value).
Three systems not recognized before as the ``S'' cases, but having
parallax errors larger than 3--$\sigma$ relative to the error median,
have been added here: UX~Eri, V1363~Eri and DY~Cet.
``N'' signifies no detection. If a system was not included in
the Hipparcos missions, then FLAG-3 is ``--''.

% Fig.2: percentage of WDS doubles + Hipparcos detections
\placefigure{visual}

In terms of its precision, the
Hipparcos sample is biased mainly by two factors, the apparent brightness
and the distance, as astrometric effects are smaller at larger distances.
As a partial result on the true frequency of triple systems, and to visualize
potential biases, we attempted to estimate the fraction
of visual and astrometric triples solely from the astrometric
data for the nearest, best observed contact binaries.
For this, we ranked the binaries according to the parallax,
with the system of rank ``1'' being the closest contact
binary 44~Boo. This sample contains 133 contact binaries brighter
than $V_{max}$ = 10 that appear in the Hipparcos catalogue and
have positive astrometric parallaxes.
In Figure~\ref{visual} the left vertical axis gives the fraction of systems
below and including a given rank, listed as visual binaries in WDS and 
suspected in Hipparcos (flags ``X'', ``G'' and ``S'').
The right vertical axis gives the total number of such systems
for the given rank.
Since one of the visual components is a contact binary, these systems
are at least triple. For the 50 closest contact systems, the
fraction of systems being at least triple shows a plateau
and appears to be 40\% although the Poissonian noise at that point
is about 9\%. Such an estimate of the triple-system
frequency, based only on the astrometric data,
is incomplete as we have a direct evidence that
the true frequency is higher: Additional 8
multiple systems with contact binaries were detected spectroscopically
(Section~\ref{spectroscopy}) without any
astrometric indications of multiplicity (not even the ``S'' flag)
in the Hipparcos astrometric data while two more very close systems
(CK~Boo and U~Peg) were discovered during the CFHT
adaptive-optics observations (Section~\ref{sec_cfht}).

\subsection{Proper-motion errors (FLAG-4)}

The Hipparcos mission provided an excellent new epoch datum for
proper motion studies of nearby stars. Proper motion data improve over
time as the time-base grows. Use of the old data led to the main
improvement in the proper motion results from the
TYCHO-2 project \citep{tycho2} when compared with
the original Hipparcos results based on 3
years of the satellite operation. We can
utilize this material the same way as described above for the
parallax errors by analyzing individual errors of the
proper motion determinations.

We note that the proper motion errors depend on many factors such as
(i)~the time interval covered by observations,
(ii)~the number and quality of astrometric data,
and (iii)~the brightness of the system.
Figure~\ref{err-proper} presents evidence that known visual systems among
contact binaries have larger proper-motion errors, so that the
situation is indeed analogous to that with the parallax errors. The
difference is in the time scale of the external orbit as
this indicator is sensitive to companions orbiting on large
orbits, with the TYCHO-2 data sensitive to
periods spanning perhaps as long as a fraction of a
century; this should be contrasted with the Hipparcos parallax errors
which were estimated from internal uncertainties of the
mission lasting 3 years.

In Figure~\ref{err-proper}, note the known members of
visual binaries such as EE~Cet or V752~Mon.
Other systems, like V376~And or V2377~Oph may
be members of triple systems with relatively long-period orbits.
For the statistics in Table~\ref{summary}, we take
as suspicious cases all systems
with proper-motion errors 3--$\sigma$ above the TYCHO-2 error median,
as before assuming $\sigma$ to be the $rms$ uncertainty of the median.
We included FU~Dra, V829~Her and V839~Oph which are close to this limit.
We note that in the error estimation,
one must be careful with systems having only 3
astrometric observations, as one erroneous observation would result in
a large error of the derived proper motion.

% Fig. 3: error of proper motions
\placefigure{err-proper}

\section{SPECTROSCOPY (FLAG-5)}
\label{spectroscopy}

\subsection{Line profiles and broadening functions}

Spectroscopy offers one of the most direct methods of discovery of
companions to contact binaries. When a companion is present,
extraction techniques
utilizing deconvolution of spectra, such as the broadening
functions (BF, see \citet{Rci2002a}) or their poor-quality
substitute, the Cross-Correlation Function (CCF), often
show one sharp peak superimposed on a background of a broad
and rapidly changing projection of the contact binary brightness
onto the radial-velocity space.

The radial velocity program at the David Dunlap Observatory
(currently at Paper~X and 90-th orbit, \citet{ddo10}) is specifically
designed to provide
radial velocity data for close binary systems with periods shorter
than one day and brighter than about 10 magnitude. It was noticed
almost from the beginning of the program in 1999 that our sample
contained many triple systems with third components frequently
dominating in light. This turned out not an accidental occurrence,
but an interesting, if mild, bias in our sample:
Our candidates have been predominantly from among recent
photometric discoveries -- frequently by the Hipparcos project --
with small photometric amplitudes. While for some of these cases
the low amplitudes happened to be due to a low orbital inclination,
for many binaries, when a companion is present,
the reason is the ``dilution'' of the photometric
signal in the combined light. These low
photometric amplitude binaries were not known until the systematic
(to $V \simeq 7.5$) Hipparcos survey took place. Spectroscopically,
these stars had been frequently overlooked as their spectra
-- even when the companion is relatively fainter -- typically
show only one sharp-line component, sometimes only with some fine
and hard to detect continuum peculiarities \citep{Rci2002a}.

At this time, out of 90 stars with new, reliable elements determined
during the DDO spectroscopic binary program, 76 are contact binaries and/or
probably contact binaries. Among those, 12 are members of visual
binaries, some already identified by other techniques.
We discovered spectroscopic companions in
V899~Her and VZ~Lib \citep{ddo04}, V401~Cyg \citep{ddo06}, as well as
VW~LMi, TV~UMi and AG~Vir (to be published). But not
all of our detections were new: Because of the moderately
poor seeing at the DDO site with a median FWHM of about 1.8 arcsec,
some already known astrometric companions were
not visible directly during the observations
and appeared first time in the spectroscopically derived broadening functions.
This way, we re-discovered several multiple systems (e.g., EF~Dra, V410~Aur).
Examples of the broadening functions for the spectroscopically
triple systems can be found in the series of DDO publications, particularly
in Paper~IV \citep{ddo04}, Paper~VI \citep{ddo06} and Paper~VIII \cite{ddo08}.
Here we show in Figure~\ref{vwlmi} the broadening functions for an even more
complex quadruple system VW~LMi whose complicated nature has been
discovered during the DDO program (to be published).

% figure with broadening functions of VW LMi
\placefigure{vwlmi}

The detection limit for spectroscopic companions, given by the ratio
of the third component light to the total light of the eclipsing
pair $\beta = l_3/(l_1 + l_2)$, at our spectral window at 5184 \AA, is
about 0.03 -- 0.04. This estimate is based on the cases of
V401~Cyg ($\beta = 0.03$) and HX~UMa ($\beta = 0.05$)
where a weak light contribution was easily detectable. This
corresponds to $\Delta m \simeq 3.8$ in the most favorable case;
normally, the detection limit depends on the
shape of the combined BF as the sharp peak
may sometimes hide within the two lobes of the binary.
An opposite extreme case of a bright companion entirely
masking any traces of the contact binary in the broadening function
is V752~Mon (new, unpublished DDO observations),
where at $\Delta m \simeq -0.5$, the companion is
a rapidly rotating A-type star with
$V \sin i \sim 220$ km~s$^{-1}$. Generally, it is difficult
to map the domain of detection of third components as it depends
not only on $\beta$, but also on the $(K_1+K_2) \sin i$ of the binary,
its mass ratio, $V \sin i$ of the third component, etc.

\subsection{Composite spectra}

While our individual broadening functions which are used for
radial velocity measurements have a limitation for the smallest
$\beta$ detectable at a given S/N, averaging of the same spectra
may show fainter companions to $\beta \simeq 0.01$
or even below.
While the binary signatures are  smeared out through averaging,
any constant component gains in definition.
This is exactly the approach taken in Paper~II
\citep{dangelo2005} where spectra previously utilized in the
DDO radial velocity have been used to detect weak TiO
features of M-type dwarfs in e.g., SW~Lac, CK~Boo and BB~Peg.
A particularly good example is SW~Lac which was not seen
in individual BF's \citep{ddo10}, but could  be
convincingly detected in the average spectrum.
The faint companions are also detectable through the
presence of the molecular bands of TiO, VO, CaH.
In the visual and near infrared regions, the TiO bands are
the dominant source of molecular absorption in stars
later than M3 (see \citet{kirk1991}). Particularly prominent
TiO features occur around 5500 \AA\ and around 7600\AA.

We refer to Paper~II \citep{dangelo2005} for
details of the spectrum averaging technique which is
discussed there with a particular stress on its biases and
limitations. The spectral region utilized at DDO for the
radial velocity program,  of about 200 -- 300 \AA\ (depending on the
CCD detector) centered at 5184 \AA,
is a very convenient one for radial velocity measurements because
the magnesium triplet ``b'' feature is strong over a wide range of
spectral types from middle A to early K types, but is not
optimal for detection of faint, low temperature companions.
For that reason, we plan to conduct a search based
on good quality, low-dispersion, red spectra
of contact binaries with a specific goal of detection of molecular
features characteristic for late-type companions.

We note that
the disentangling technique \citep{hadrava2004}) has a potential
of determination of the individual spectra from multi-component
spectra undergoing radial-velocity shifts. The method produces
orbital parameters of the spectroscopic binary as well as
information on any additional spectrum in the system; thus
telluric lines as well as lines of a third component are naturally
determined.

FLAG-5 in Table~\ref{summary}
gives information on triplicity as determined from
spectroscopy, both in the radial velocity program
and in the averaged spectral data. The codes used are:
``D'' used for detection, ``S'' for suspicion,
``N'' for not detected and ``--'' not yet observed or not
analyzed.

\subsection{Systemic velocity changes}

A scatter or trends in the center-of-mass velocity, $V_0$,
may indicate presence of companions. Because
radial velocities usually come from different sources with different
systematic errors and because contact binaries show large
broadening of spectral lines, this is a difficult-to-apply
and generally poor indicator of
multiplicity. As we will see below, the required accuracy for $V_0$
(i.e.\ precision combined with systematic uncertainties) for the
well known close visual multiple systems of 44~Boo and VW~Cep is at a
level of 1~km~s$^{-1}$, which is very difficult to obtain for
contact binaries. To complicate things, very few contact binaries
have been observed for radial velocities more than at one epoch so that
a comparison of published systemic velocities can be done for only
a few objects:
\begin{itemize}
\item AB~And has three sets of spectroscopic elements
      \citep{struve1950,hrivna1988,pych2004}. While the second and
      the third source give similar $V_0$, the first gives a velocity
      shifted by $-16$ km~s$^{-1}$.
\item 44~Boo has several RV studies. The results of early studies are
      discussed by \citet{hill1989}, with the most reliable data
      subsequently published
      by \citet{ddo04}. Systemic velocity changes were observed to range
      between $+3.9$ and $-32$ km~s$^{-1}$,
      which is much more than the expected full amplitude for the
      well studied visual system of only 0.7 km~s$^{-1}$. We suspect that
      the scatter in $V_0$ is purely instrumental in this case.
\item VW~Cep was observed 5 times; see \citet{prib2000} for the discussion
      and references. The systemic velocity shows a much larger spread
      ($-35$ to $+10$ km~s$^{-1}$) than
      can be expected from the known third body, expected to cause a full
      amplitude of $\approx$ 5.8 km~s$^{-1}$. Again, the most likely reason
      for the scatter is instrumental.
\item Spectral lines of a third component were detected in SW~Lac
      by \citet{HenryMki1998}.
      Three available systemic velocities \citep{struve1949,zhailu1989,ddo10}
      range between $-15$ and $-10$ km~s$^{-1}$,
      which can be considered as a case of an
      exceptionally good agreement and thus constancy
      of the mean velocity. This is somewhat paradoxical as we know that
      the system contains at least one additional component.
\item V502~Oph has four studies
      \citep{strugrat1948,struzeb1959,kinghild1984,pych2004}.
      \citet{struzeb1959} noted a $-13$ km~s$^{-1}$
      shift in radial velocities between the two of their runs.
      Lines of a third component were detected by \citet{HenryMki1998}.
\item There exist seven published radial velocity datasets for
      W~UMa \citep{Rci1993a}. Systemic velocities show large variability
      in the range from $-43$ to $0$ km~s$^{-1}$. Part of the scatter
      can be instrumental, but presence of a third
      body in the system is possible.
\item Possible multiplicity of AW~UMa has been discussed in \citet{prib1999b}.
      The observed systemic velocity changes would require a relatively
      massive body which was not detected in BF's of the system \citep{Rci1992}.
      The system is a prime candidate for a new spectroscopic study.
\end{itemize}

On the basis of the observed large systemic velocity changes, two systems,
W~UMa and AW~UMa, are coded as ``S'' (suspected) by the FLAG-5 in
Table~\ref{summary}.

\section{INDIRECT MULTIPLICITY INDICATORS}
\label{indicators}

\subsection{The light-time effect ``LITE'' (FLAG-6)}

Accelerated or delayed times of eclipses in close binaries are normally
interpreted as a result of apparent period changes.
These changes may result from the binary revolution
on a wide orbit around a center of mass with a companion; this
is then frequently called the ``LITE'' for the LIght-Time Effect
\citep{borhe1996}.
Because intrinsic period changes in close binaries
are quite common due to their strong physical interaction,
the presence of the LITE is usually not conclusive, but can
prompt other observations. Such was the case of VW~Cep where
periodic variation of the eclipse times was interpreted by the
LITE very early by \citet{payne1941}.
In 1974, the third component had been indicated by the transverse motion of
the contact pair and then was directly observed \citep{Hershey1975}.

The period of an eclipsing binary is the most precisely
determined orbital element due to accumulation of period deviations
over many orbits. Traditionally, timing of eclipses
has been a separate activity
concentrated on the light minima only \citep{kreiner2001},
but new techniques utilize whole light curves \citep{gensmi2004}
while new light-curve fitting codes even include the LITE
to enable the analysis of short-period triple systems
\citep{hamme2005}.

The LITE technique has several advantages over the other approaches
because (i)~it is distance independent, (ii)~light minima can be
observed with small
telescopes, (iii)~the time delay/advance effect, in contrast to its
spectroscopic-amplitude counterpart (the variation in $V_0$),
increases with the period of the third body,
and (iv)~detection is independent
of the relative brightness of the third component.
The technique has also pitfalls:
(i)~The behavior of (O--C) residuals (observed minus calculated
time of eclipse) very often indicates periods being
similar to about 2/3 of the used time interval; this may be a signal
of a failure in the main LITE assumptions leading to over-interpretation
of the data;
(ii)~Detection of short-period orbits may be negatively influenced by
the enhanced surface activity \citep{prib2005} and
migrating spots can sometimes cause wavelike behavior of the (O--C)
diagrams (see e.g, \citet{kalimer2002}); fortunately,
the spot-induced shifts of the eclipse minima
usually only increase the scatter of observed minima times
\citep{prib2001};
(iii)~Another possible interpretation of the
cyclic period changes is a periodic transfer of the orbital angular
momentum to magnetic momentum in active systems \citep{apple1992};
while this may be the case, the actual confirmation of
this mechanism still remains to be done and is still controversial
\citep{prib2001,afsar2004}.

The formulae for the theoretical LITE computation can be
found in \cite{Irwin1959}. An
observed time of the minimum light can be predicted as follows:
\begin{equation}
JD(Min I) = JD_0 + PE + QE^2 + \frac{a_{12} \sin i}{c}
\left[\frac{1-e^2}{1+ e \cos \nu} \sin(\nu + \omega) + e \sin
\omega\right],
\end{equation}
\noindent where $c$ is speed of light,
$JD_0 + PE + QE^2$ is the quadratic ephemeris of
eclipsing pair (a secular change of the period measured by $Q$ is the most
likely type of change for the binary itself), $a_{12}$, $i$, $e$, $\omega$
are the orbital parameters and $\nu$ is
the true anomaly of the binary on the orbit around the common center of
gravity with the third body. Since the LITE operates in the radial
direction through a mechanism similar to the Doppler effect, the
LITE parameters are similar to spectroscopic elements.
The problem of cyclic variations in the period is one of a simple
model fitting: Given pairs $JD(Min I)$ and $E$, one must
optimize parameters $P_3, a_{12} \sin i, \omega, e, T_0$ and ephemeris
of the eclipsing pair $JD_0, P, Q$.

For the present analysis, the contact binaries were selected from the updated
version of the Catalogue of Contact Binary Stars (hereafter CCBS,
\citet{prib2003}) currently containing 391 systems. Among these,
129 have more than 15 CCD or {\it pe\/} minima in the Cracow database
(the May 2004 version, \citet{kreiner2001}).
Because standard errors are available for only a small part of the minima
times and cannot be trusted, we applied an arbitrary weighting
scheme with the weights assigned as $w$ = 10 to {\it pe\/}, $w$ = 5 to CCD,
$w$ = 3 to visual and photographic and $w$ =1 to
``patrol'' and single-epoch plate minima. The weight $w \le 3$
minima were used only in cases when no other minima were available.
Only 18 out of 129 systems provided stable
LITE solutions. In our sample of contact binaries brighter than
$V_{max} = 10$ there are 57 systems with a sufficient number of minima;
among them 11 have feasible LITE solutions.

%Fig.4a O-C for 2 systems
\placefigure{omc}

An analysis of all available minima and new LITE solutions
are presented here for TY~Boo, TZ~Boo, CK~Boo, GW~Cep, UX~Eri,
V566~Oph and BB~Peg. The resulting parameters are given in
Table~\ref{lite} while Figure~\ref{omc} illustrates two
rather typical cases of TZ~Boo and BB~Peg.
For SW~Lac and CK~Boo, the orbital
period $P_3$ is very uncertain and its inclusion as an unknown
in fact destabilized the solutions. In case of UX~Eri,
inclusion of the orbital eccentricity destabilizes the solution.

In addition to the approximate orbital elements, Table~\ref{lite}
gives minimum mass of the third body determined from the mass function
and total masses of the eclipsing pair. For the latter, we used
the CCBS data, in available cases modified by values determined in the DDO
program of spectroscopic orbit determinations
(see \citet{ddo10}). We also give
the expected angular separation of the third (or multiple components)
in arcsec for cases of available trigonometric parallaxes, evaluated as
$\alpha = \pi a_{12} \sin i_3 (m_{12} + m_3)/m_3$.

%Table 3:  LITE
\placetable{lite}

Inspection of Table~\ref{lite} shows that minimum masses for some systems
(AH~Vir, ER~Ori, CK~Boo and the outer body in SW~Lac)
appear to be unrealistically high which
indicates that cyclic variations may have been caused by other
effects, perhaps intrinsic to the binary. Predicted mean
separations for several systems (e.g., AB~And, CK~Boo, V566~Oph) are
accessible to speckle or direct detection. In fact, CK~Boo was
independently detected as a visual double during CFHT observations
(Section~\ref{sec_cfht}) while
an UV excess may indicate existence of a white dwarf in
the system \citep{krez1991}.

SW~Lac
requires a special attention as this may be a quadruple system. The first
consistent LITE solution \citep{prib1999a} was not stable and could not
fully explain the behavior of older minima. Multiplicity of the
system was clearly seen in the spectra of \citet{HenryMki1998}. We have
found it to be well separated in the CFHT $K$ and $H$-band
Adaptive Optics observations (Section~\ref{sec_cfht}) with the third
component too far on the sky to cause the observed LITE.

Detection of the LITE obviously depends on the number of
available minima, although other factors, such as their temporal
distribution, are also important. Thus, systems can be ranked according
to this number which reflects our chances to detect a cyclic period change.
In Fig.~\ref{litefig} the fraction of LITE-detected multiple systems
is plotted with respect to the rank of system, sorted according to the
number of the available minima times. It is obvious that detection of multiple
components among the studied systems is strongly dependent on the
availability of data. For the first best observed 20 systems, the multiplicity
rate appears to be as high as about 60\% $\pm$ 17\%. It should be
noted that most of those systems are the Northern hemisphere ones as
too few eclipses have been observed for Southern contact binaries to
permit LITE studies. We will return to the point
of the disparity between the hemispheres in the conclusions
of the paper (Section~\ref{future}).

% Fig.~4b LITE
\placefigure{litefig}

The chance of detecting a third body by the LITE
is restricted to periods from a
few years (where the (O--C) amplitudes are small) to
several tens of years by the available span of the
data; indeed, most systems in Table~\ref{lite} have
periods about 30 years. If the third component is the
dominant source of light, it can reduce the
photometric amplitude of the eclipsing pair
(e.g., the case of TU~UMi, \citet{pychrci2004})
making determination of the eclipse moments very imprecise or even
entirely impossible. In view of this observational bias, the
60\% incidence of triple systems may be considered as possibly a
low limit to an even higher value.

In Table~\ref{summary} the resulting LITE status
(FLAG-6) is coded as follows: ``D''
for a stable solution, ``S'' for not fully stable solution or/and
less than 1.5 cycles of $P_3$ covered, ``N'' for no detection, and
``--'' for too few minima or less than 10 years of CCD or {\it pe\/}
observations. In our view, the LITE is almost certainly present in AB~And,
VW~Cep, AK~Her, SW~Lac, XY~Leo, and AH~Vir.

\subsection{The period--color relation and the color
index peculiarity (FLAG-7)}

The period -- color relation for contact binary stars was discovered
observationally by \citet{Eggen1961,Eggen1967} and was utilized
later for various theoretical investigations of these systems.
The relation is a consequence of contact binaries being
Main Sequence objects, with all implied correlations between the mass,
effective temperature and radius, hence the size and thus the period.
Observationally, the relation is a very useful tool
in recognizing contact binaries
in deep variability surveys and, together with the luminosity calibration
$M_V = M_V(\log P, B-V)$ \citep{Rci1994,Rci1998,Rci2002c},
it forms a powerful discriminator for identifying contact
binaries among short-period variable
stars. The color index $B-V$ can be replaced by
$V-I$ or any other index; an unpublished $V-K$ calibration also
exists (Rucinski, in preparation).

As pointed before \citep{Rci1997,Rci1998,Rci2002b}
the period -- color relation must have a
short-period, blue envelope (SPBE). Its locus corresponds to the Zero Age
Main Sequence of least evolved stars, when it is applied to stars
in contact and expressed in terms of the period and a color index.
Evolution and expansion of components can lead to
lengthening of the period and to reddening of the
color index; reddening may be also caused by the interstellar extinction.
Thus, one does not expect contact binaries to appear blue-ward of
the SPBE. Yet, some close binaries do show up there.

\citet{Rci1998} suggested a simple approximation to the
SPBE, $(B-V) = 0.04 \times P^{-2.25}$,  and an explanation
for the blue outliers as non-contact stars, possibly Algols
or $\beta$~Lyr binaries with a blue component dominating in the
combined color index; a lowered metallicity may also explain some
of the blue deviations \citep{Rci2002c}.
However, some of the blue outliers may be multiple systems with
the color index peculiarity caused by a blue third star.

\placefigure{spbe}

The period-color relation for 151 contact binaries brighter than
$V_{max}$ = 10 with the $B-V$ data
from TYCHO-2 is shown in Figure~\ref{spbe}.
One can immediately see that several
systems are substantially bluer than the SPBE. Some
of them are indeed known triple systems (e.g., V867~Ara, KP~Peg).
The most deviating systems, and the strongest cases to suspect
presence of an early-type component, are V445~Cep and V758~Cen.

We are very much aware of a possibility of over-interpreting
the blue deviations in Fig.~\ref{spbe} so we do not claim any detections
here and point only some systems as suspected cases of multiplicity.
This is marked by the flag ``S'' in Table~\ref{summary}
for cases when the color index deviates by more than
1--$\sigma$ from the SPBE line. Systems not included in the TYCHO-2
are given the ``--'' flag. ``N'' signifies no detection.

\subsection{X-ray emission (FLAG-8)}

Contact binary stars consist of solar-type stars spun into very
rapid rotation by the tidal forces. As a consequence, they are
very active and show elevated chromospheric and coronal activity.
An extensive body of data exists on their X-ray emission
\citep{Stepien2001} and, in particular, on the correlation of the
X-ray coronal emission with the two most relevant parameters, the
effective temperature, $T_{eff}$, and the orbital period, $P$; the
former controls the depth of the convective zone while the latter
controls the amount of vorticity available for generation of
magnetic fields. A convenient distance independent quantity is the
ratio of the apparent X-ray and bolometric fluxes, $f_X/f_{bol}$.
It expresses the X-ray ``efficiency'' of how much of the total
available energy is channeled into the X-rays. $f_{bol}$ can be
estimated from the visual magnitude of the star utilizing the
bolometric correction $BC(T_{eff})$. Unfortunately, many fainter
binaries still do not have any standardized (say $UBV$)
photometric data to estimate even an approximate value of
$T_{eff}$. Therefore, to use a homogeneous and simple treatment,
we used the shifted SPBE relation (by subtracting 0.2 mag) to
estimate an approximate value of the $B-V$ color index from the
orbital period (see Figure~\ref{spbe}); because the bolometric
correction changes very little between A0 and G5 spectral types,
and we are looking for an order-of-magnitude effects in
$f_X/f_{bol}$, the resulting approximation is adequate. The X-ray
fluxes were estimated using the procedure described in
\citet{Stepien2001} from the observed RASS X-ray count rate and
the hardness ratio (between the bands 0.1 -- 0.4 keV and 0.5 --
2.0 keV). The measured X-ray fluxes were corrected for the
background count rate. In the computation of bolometric fluxes we
took $2.5 \times 10^5$ erg~cm$^{-2}$~s$^{-1}$ as a flux for
$M_{bol}$ = 0. The bolometric corrections were interpolated in the
table given by \citet{popper1980}.

We utilized the RASS Catalog \citep{voges1999,voges2000} to find
$f_X$ in the band 0.1 -- 2.4 keV  for all contact binaries of our
sample. Although the catalogue contains its own
cross-identifications, after realizing many omissions, we
cross-correlated the RASS catalogue data with the IRSC2
coordinates taken from SIMBAD for all CCBS systems. The angular
separation of 36 arcsec was taken as a limit for detection. Six
additional distant sources identified in SIMBAD have been also
included. Although not all detections are secure due to the
relatively low pointing precision (typically 10 -- 20 arcsec),
most identifications are unambiguous as most of the targets do not
have any other objects within 0.5 degree. Some binaries of our
sample were not detected providing an useful upper limit on
$f_X/f_{bol}$, derived from the assumed detection threshold of
0.01 counts/sec. To better define the lower envelope of the upper
limits on $f_X/f_{bol}$, we added also systems fainter than
$V_{max}$ = 10mag.\ from CCBS. The results are shown in
Figure~\ref{Xrays}.

\placefigure{Xrays}
% Fig 6: X-rays

Following the previous investigations \citep{CrudDup1984,Stepien2001},
we expected the ratio $f_X/f_{bol}$ to show a weak
dependence on the binary orbital period close to a saturation limit
at $f_X/f_{bol} \simeq 10^{-3}$. However, the scatter in Figure~\ref{Xrays}
is large with some deviations reaching factors of 10 or 100 times from
the average. We argue that companions to contact binaries may
be the cause of these large deviations and that the
deviations may go both ways:
(i)~When an early-type contact binary has an
M-dwarf companion or -- particularly -- a {\it binary\/} M-dwarf
companion (as in the case of XY~Leo) then $f_X/f_{bol}$
can be strongly elevated, but (ii)~when the
contact binary is of solar or later spectral type, its
magnetically inactive
early-type companion may reduce the value of
$f_X/f_{bol}$ (the cases of 44~Boo or V752~Mon). Indeed,
this is exactly what we do see in Figure~\ref{Xrays} for the known triple
systems: (i)~In the left lower corner, a few system with the relatively
low flux ratio are in majority triple systems where the X-ray flux is
diluted by the dominant early-type, single star (e.g., V345~Gem);
(ii)~Majority of short-period systems with high relative
X-ray fluxes are triple systems; in some cases a late-type, binary,
BY~Dra-type companion provides most of
the X-ray flux (e.g., XY~Leo), and
(iii)~Several early-type, long-period contact
systems, expected to be practically inactive,
show enhanced X-ray flux (V335~Peg, DO~Cha or RS~Col) which may
indicate that they host a late-type, active companion.

The $f_X/f_{bol}$ ratio should be used as an indicator of
multiplicity with a considerable caution,
particularly in view of the large number of upper limits to the X-ray
flux values.
However, it may be highly indicative in individual cases already
signaled by other techniques. In Figure~\ref{Xrays} we marked a line
delineating the low X-ray luminosity envelope (LXLE) of the observed
$f_X/f_{bol}$ ratios. The line was selected rather arbitrarily to
encompass most of the systems.
Systems with $f_X/f_{bol}$ hundred times above
the LXLE are marked in Table~\ref{summary} by ``L'',
for a ``Late-type companion''. Similarly late-type contact
systems with an abnormally low emission, those below the LXLE,
are marked by ``E'', for an ``Early-type companion''.
For periods longer than 0.5 days, the LXLE is not defined because such
contact binaries of spectral types early-F or earlier are not expected
to show any detectable X-ray radiation.

\subsection{Lunar occultations}

The high speed photometry of lunar occultations can lead to
detection of multiple components and the technique is
widely used \citep{Richichi2002,kazan2002}.
An angular detection limit for bright binaries,
given by the Moon-limb diffraction angular scale is about
2 mas \citep{warn1988}. For faint stars, this limit is modified by
the diameter of the telescope through the photon noise
in very short integrations utilized in the high-speed photometry.
Due to the apsidal motion of its orbit, the Moon can
occult about 9\% of the sky. From among 391
contact binary stars in the electronic version of
CCBS \citep{prib2003}, 26 lie within the path of the
Moon ($b = \pm 5^\circ 9'$ around the ecliptic).

Among systems in Table~\ref{summary}, the best candidates for
occultations are bright systems AQ~Psc ($V=8.68$,
$b=-0.867^\circ$), V4408~Sgr ($V=8.29$, $b=+2.907^\circ$),
V781~Tau ($V=8.90$, $b=+3.541^\circ$). Somewhat fainter, but
interesting systems are: CX~Vir, V1123~Tau, AP~Leo, AM~Leo,
UZ~Leo, XY~Leo, RZ~Tau, TX~Cnc, VZ~Lib, AH~Aur and CT~Tau. The
binaries which can be studied using the Moon occultation technique
are marked by ``M'' in the column ``Moon'' in
Table~\ref{summary}). So far only the quadruple system XY~Leo was
observed \citep{evans1986}, but with a negative result.

\subsection{The third light in light curve solutions}

The light curve analysis sometimes indicates a need of a ``third
light'' as an additive term in the overall brightness
budget. A third component to a binary system can be such
third light, but light curves are described by so many
parameters, that this is frequently an unreliable indicator of
multiplicity. Certainly, for light curves of partially
eclipsing spherical stars in only one band,
the third light is just an ``easy escape'' to obtain a
solution, but for contact binaries its
physical reality becomes even more dubious
because a more complex geometry.

Light-curve synthesis solutions
indicating a need of a third light usually show a tight correlation of the
third light with the inclination angle and the mass ratio.
Even for best constrained  solutions of contact binaries
showing total eclipses, where inner eclipse contacts can be
well defined and the mass ratio can be determined solely
from light curves \citep{Mkidoug72}, a correlation between mass ratio
and third light remains. An example is the case of EF~Dra,
where the light curve indicated $q_{phot} = 0.10$ \citep{robsca1989},
while spectroscopic mass ratio is $q_{spec} = 0.16$ \citep{LuRci1999}.
Usually without spectroscopic determination of the mass ratio
it is hardly possible to detect third light from light curves
with the usual photometric precision of $\approx 0.01$ mag.
The situation improves when the third component has a markedly
different spectral type than the binary and when photometry is available
in several bandpasses; then -- instead of uncorrelated values
of $L_3$ in each band -- the proper approach is is to use
the radius and temperature of the third star. At this time,
we feel that we have no clear case of a third-light in any
of contact binaries of our sample.

\subsection{Precession of the orbital plane}

A close and massive third body orbiting a binary system can cause
precession of the orbital plane and an apsidal motion of the close pair.
The largest dynamical effects are expected for contact binaries
which contain least angular momentum among all binaries.
Precession of the orbital plane in an eclipsing binary will
result in changes of its orbital inclination and thus
changes of the photometric amplitude of the eclipses.
The practical use of the method consists of a
comparison of high-quality light curves obtained over long
time intervals in search of amplitude variations.

The precession period depends on the period ratio $P_3^2/P_{12}$ and
(weakly) on the ratio of the outer orbit angular momentum to the
total invariable angular momentum and on the ratio of the total
triple system mass and to that of he third body mass. The nodal period
is, to the first order (see Eq.(27) in \citet{Sod1975}), given by:
\begin{equation}
P_{node} = \frac{4}{3}\frac{m_{123}}{m_3}
\frac{P^2_3}{P_{12}}(1-e_3^2)^{3/2}\frac{L_{3}}{L} \sec j,
\end{equation}
\noindent
where $m_{123}$ is the total mass, $L$ is the total invariable angular
momentum of the triple system, $L_{3}$ is the angular momentum of
the outer orbit and $j$ is the relative inclination of the orbits.

Systematic changes in the eclipse depth indicating orbital
precession have been observed in several detached binaries
\citep{Mayer2004}. However, none of contact binaries has
shown this effect, although many are observed
on a routine basis. Chances to see the effect
are high due to (i)~a very low angular momentum of a
typical contact binary, (ii)~its short orbital period, and -- usually --
(iii)~a low total mass of the eclipsing pair. Any adverse effects
of the photometric system mismatch are also minimized thanks to
the same temperature of both components so that the light-curve
amplitude is almost independent of the wavelength used, permitting
comparison of curves obtained in different photometric systems.

\section{STATISTICS AND RESULTS}
\label{conclude}

We have carefully inspected our list of contact binaries brighter
than $V_{max} = 10$ with periods shorter than one day to
establish all cases which appear to indicate multiplicity.
The last column of Table~\ref{summary}
contains the letter code ``Y'' for cases of sufficient
information to claim that a contact binary is indeed
part of a multiple system. Our decision was based on a combined
``weight'', as determined from individual detection techniques,
equal or exceeding the value of 1.0.
In adding the weights we used a weight of 1.0 for individual
unquestionable detections and a weight of 0.5 for suspected cases.
Although this approach produces non-subjective and repeatable
results, we feel that not all cases for accepted detections
are equally valid. We comment on individual cases requiring explanation
in Table~\ref{notes}.

We see 64 triple systems among 151 objects of our sample, giving the
nominal lower limit to the frequency of triple systems of 42\% $\pm$ 5\%.
This result is for both hemispheres, which have not been
equally well surveyed to apply all available methods, with the
Southern hemisphere contributing very few binaries to the final
statistics. Indeed, if not for the Hipparcos mission which
discovered a few visual binaries containing
contact binaries (e.g., V867~Ara, CN~Hyi),
there would be almost no known visual binary detections
in the Southern sky. While on the Northern sky
($\delta >0$), using all available techniques,
we see 52 triple systems among 88 objects,
giving the frequency of 59\% $\pm$ 8\%, the Southern sky
yields only 12 systems out of 63, corresponding to
19\% $\pm$ 6\%. The reason is the low
number of contact binaries discovered so far in the South,
a meager volume of photometric data even for the known ones
(some binaries were observed the last time
half a century ago!), a lack of any spectroscopic support and
a short time base for techniques requiring accumulation over
time (such as visual orbits, proper motions and LITE).

In principle, our result on the low limit to the
frequency of triple systems containing contact binaries
of 59\% $\pm$ 8\% should be compared with control
samples for single stars and for wider binaries. Such samples
are only now being created with the particularly strong
contributions of \citet{Tok1997,Tok2004}. However,
the data are fragmentary and -- even for bright,
nearby, single stars --  heavily biased by the distance
and brightness selection effects.
\citet{Tok2004} found that among 18 closest dwarfs to 8 pc, 5 exist
in multiple systems, giving the frequency of 28\%.
However, a deeper search based on 3383 dwarfs to 50 pc \citet{Tok2004}
yielded only 76 multiple systems, 2.2\%, with the
heavy bias against discovery setting in for systems beyond 10 pc.
While our final conclusion is that triple systems are
very frequent for the contact binaries, we note that
several of our techniques are different from those of
Tokovinin and are distance
independent; this may explain our high result when compared with
that of \citet{Tok2004}, in spite of the fact that
our targets are spread in distance from 13 pc
to hundreds of parsecs\footnote{The distance to closest
contact binary 44~Boo is 12.77 pc.}. The situation appears to be
different for binaries. \citet{Tok1997} found that the frequency of
triplicity and multiplicity among spectroscopic binary
stars with orbital periods shorter than 10 days
appears to be high: As many as
43\% $\pm$ 8\% (26 out of 61) among nearby (within 100 pc),
low-mass (0.5 to 1.5 M$_\odot$), spectroscopic binaries
cataloged in \citet{batten1989} have known tertiary components.
Our estimate for the Northern hemisphere contact binaries
of 59\% $\pm$ 8\% is consistent with Tokovinin's results
within the statistical errors, but is slightly higher.
This may indicate that the frequency keeps on increasing
for shorter periods, but that the underlying physical
causes for periods $P < 10$ days may be similar.

We note that the main route for detection of
companions is the small angular separation
on the sky. Among the 64 positive detections, 46 systems are
known visual, previously cataloged
WDS and Hipparcos binaries or new discoveries of the
new CFHT, adaptive-optics IR observations; the
latter nicely complemented the extant data
for M-dwarf companions at angular separations of
about 0.1 arcsec to a few arcsec. In turn, most of the spectroscopic
detections have been of systems already known as visual
binaries; spectroscopy can, however, help in determining
the physical characteristics of the companion and firm up
individual cases.

The visual technique, combined with the parallax data permits to
evaluate projected physical separations between the companions.
The projected separations are moderate and are distributed in the range
between 6 and $10^4$ AU, or $2.9 \times 10^{-5}$ pc and 0.049 pc
(Table~\ref{summary})\footnote{We exclude EM~Cep and AH~Cnc
which are members of open cluster and whose projected separations
are 49,700 AU and
24,000 AU. We do not consider these systems as multiple ones.}.
Figure~\ref{separ} shows a histogram of the projected
separations. The distribution may partly reflect
observational biases, but it does
show that the separations are much smaller than typical distances
between stars in the solar neighborhood of about one
parsec. Thus, even the widest pairs with
angular separations of several tens of arcsec can be regarded
as gravitationally bound; an example is AW~UMa where
for the angular separation of 67 arcsec, the projected
separation is 4,400 AU.
But -- physically -- can these multiples be evolutionary connected
at such disparate separations ranging between stellar sizes and
thousands of AU's? Can the Kozai cycle work at such large distances
or do we see now only the results of it acting well in the past?
Answers to these exciting questions are beyond the scope of
this paper.

%figure with histogram of separations \label{separ}
\placefigure{separ}

\section{FUTURE WORK AND CONCLUSIONS}
\label{future}

The approach that used is this paper was a straightforward one: We
simply counted contact binaries which appeared to have
companions. We totally disregarded the important
matter of {\it what are the companions and what would be the limitations in
their detection\/}. Some classes of stars, such as white dwarfs or neutron
stars, would require utilization of special techniques; some
classes may be entirely undetectable. We leave these issues
open for future investigations of this series.

The use of several different observational techniques was
certainly useful as none gives an unquestionable proof of
multiplicity and each has its own biases and limitations. We used
too many techniques to analyze selection effects of each; such
in-depth studies of biases for each technique would inflate the
study and would probably not be warranted at this preliminary
stage. But we know that biases are important: Taking the
astrometric data for the 50 best observed systems in the whole sky
brighter than $V_{max}=10$ (Section~\ref{direct},
Figure~\ref{visual}), we see a frequency of 40\% $\pm$ 9\%, but
this number would rise to 46\% $\pm$ 10\% if we disregarded the
magnitude limit and argued that objects best observed
astrometrically would provide an even more reliable sample 
(indeed, there are many bright systems with poor data). 
The sky location is also
important: for the Northern sub-sample of astrometrically well
observed stars to $V_{max}=10$, the frequency increases from
40\% $\pm$ 9\% to 50\% $\pm$ 10\%. 
Similarly, entirely independent indications for the 20
best observed systems for the LITE suggest 60\% $\pm$ 17\%; among
those LITE solutions, among these only two are for nominally
Southern systems, although both were observed from the North
(ER~Ori with $\delta = -8.5^\circ$ and UX~Eri with $\delta =
-6.9^\circ$).

The most obvious observational bias caused by the brightness
of the system is demonstrated in Fig.~\ref{multi}, where the fraction
of detections is plotted for all systems brighter than the given
magnitude rank. While for systems brighter than $V_{max}$ = 9 mag
we see an almost constant fraction of multiples, there
is a small but definite decrease of detections for fainter systems.

In addition to in-depth analyzes of selection effects and biases
of individual techniques, we issue the usual plea to the
observers: More data are needed! What is particularly needed
are photometric data in standard systems. If not for the Hipparcos
mission, which led to many discoveries of contact binaries
and if not for the photometric part of the
TYCHO project, many contact binaries would be unknown or without
most elementary data on color indices. All contact systems
require also radial velocity spectroscopic support.
The situation with
the Southern sky is particularly grave. There exists no
spectroscopic survey similar to that being conducted now for
short-period Northern binaries at DDO so that even relatively
bright southern contact binaries (TY~Men, DE~Oct, MW~Pav, CN~Hyi,
WY~Hor, V386~Pav, etc.) still await spectroscopy or even
conventional precision photometry. Very few Southern
binaries are followed over
time to provide LITE information. For many binaries, the
last datum on the moment of minimum is the Hipparcos
$T_0$\footnote{This remark unfortunately
applies also to some Northern binaries which
do not happen to be popular...}.  Interesting
Southern systems omitted by Hipparcos (e.g., BR~Mus,
TV~Mus or ST~Ind) frequently have the last minima obtained
a quarter or half a century ago. We point also that the
technique of lunar occultations still remains to be
fully utilized as it can provide angular resolution
of about 2 mas and thus can be a powerful tool for the detection of
close visual companions.

We conclude with a summary of our results: Our best estimate of the
lower limit to the triple star incidence among
contact binaries appears to be 59 $\pm$ 8\%,
as based on the Northern sky to $V_{max}=10$. We know that
this limit may raise as suspected cases are analyzed more thoroughly.
If, in addition to unquestionable detection of
64 multiple systems among 151
targets over the whole sky, all suspected 24
cases in Table~\ref{summary} were included as detections,
then the frequency for the whole sky would increase from
42\% $\pm$ 5\% to 56\% $\pm$ 6\%, while that for the Northern
hemisphere alone it would reach 72\% $\pm$ 9\%.
Our results are consistent with those for the
magnitude-complete Hipparcos sample to $V_{max}=7.5$
\citep{Rci2002b},
with 17 multiples among 35 systems ($48 \pm 12$\%), although for
that sample a large fraction of all systems were
newly discovered eclipsing binaries with currently
incomplete data. Remembering that we could evaluate only
a lower limit to the frequency of triple systems,
and that some multiple systems are beyond
reach of any current technique of detection, this number
is not far from one indicating a possibility that
all contact binaries originated in multiple systems.

\acknowledgements

Thanks are expressed to the anonymous reviewer who very carefully
read the first version of the paper and suggested very useful
changes to it.

Support from the Natural Sciences and Engineering Council of Canada
to SMR is acknowledged with gratitude. The travel of TP to
Canada has been supported by a IAU Commission 46 travel grant and a
Slovak Academy of Sciences VEGA grant 4014.
TP appreciates the hospitality and support of the local staff
during his stay at DDO.

The research made use of the SIMBAD database, operated at the CDS,
Strasbourg, France and accessible through the Canadian Astronomy
Data Centre, which is operated by the Herzberg Institute of
Astrophysics, National Research Council of Canada. This research
made also use of the Washington Double Star (WDS) Catalog
maintained at the U.S. Naval Observatory.

\clearpage

\noindent
Captions to figures:

\bigskip

\figcaption[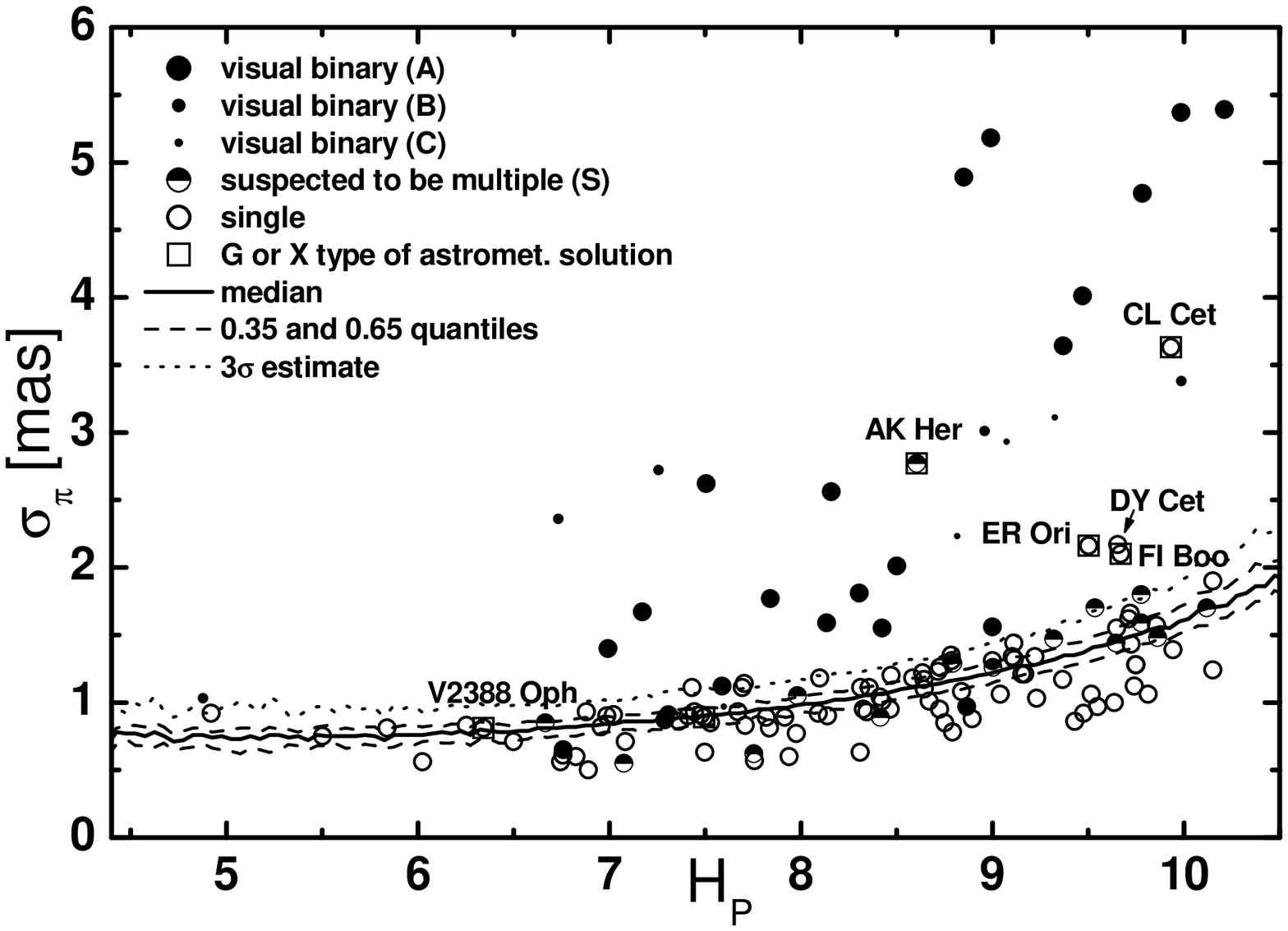]
%\figcaption[corr-mag.eps]
{\label{errpx-V} The parallax error
versus the median $H_p$ magnitude for contact binaries (circles).
The same dependence for single stars in the Hipparcos catalog is shown
by a continuous line (labelled ``median'') in the lower part of
the figure with the corresponding statistical uncertainty $\pm
1$-$\sigma$ limits (labelled ``0.35 and 0.65 quantiles'') given by
the broken lines and with the positive  3--$\sigma$ range given by
the dotted line. The known visual multiple systems among contact
binaries are marked by filled circles, with their sizes indicating
the quality of the astrometric solutions, as given in Catalog H61
field (A - ``good'', B - ``fair'', C - ``poor''). Binaries with
``G'' or ``X'' type astrometric solutions (see the text) are
additionally in squares. We argue that binaries above the curve
are strongly suspected to belong to multiple systems. }

\figcaption[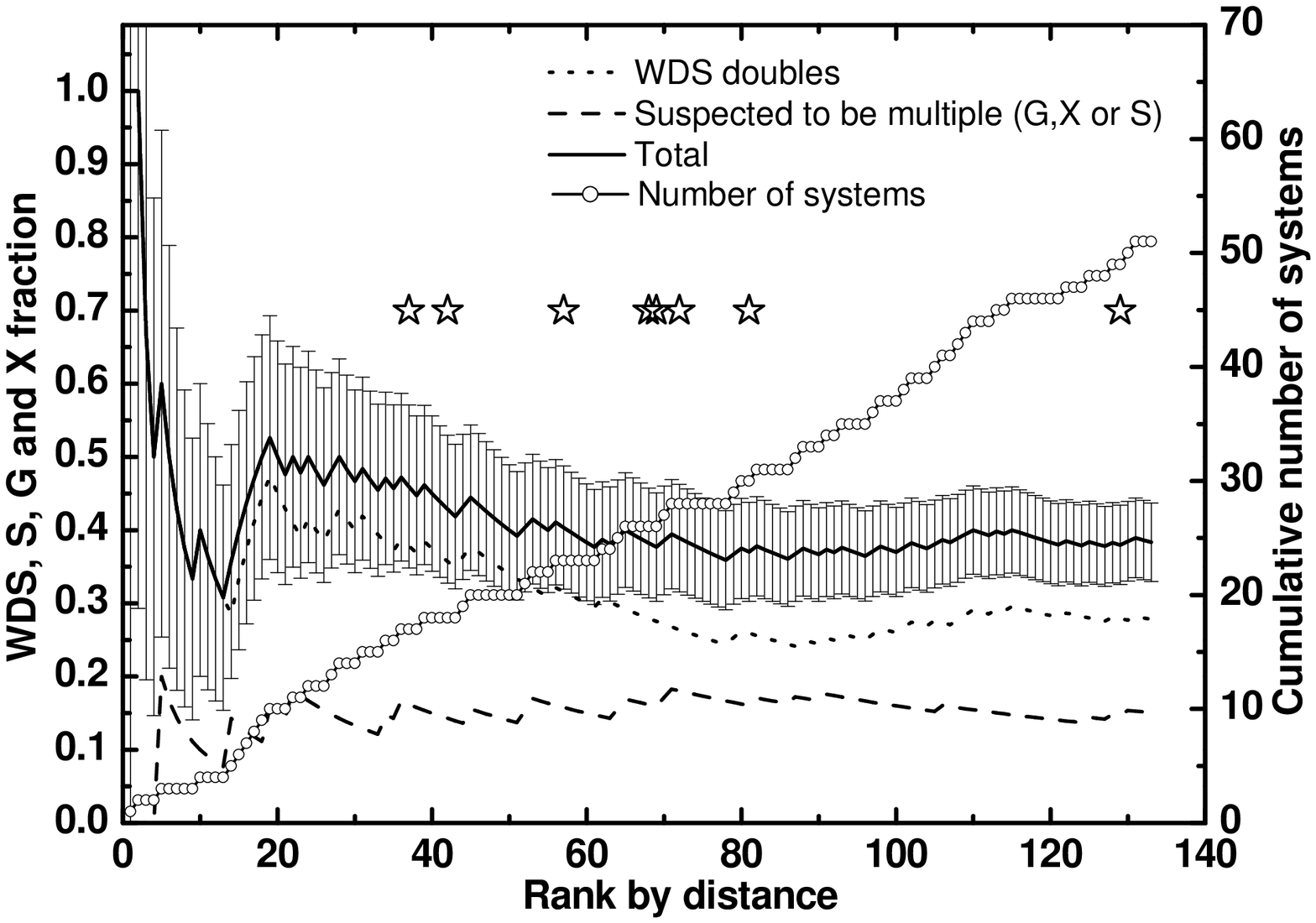]
%\figcaption[visual.eps]
{\label{visual} The fraction of visual double systems 
among the Hipparcos contact binaries,
ranked with the increased distance for the sample to $V_{max} = 10$.
The fraction for systems suspected of being double stars
(``S'' flag in the field H61) or showing an acceleration
term or a stochastic solution (``G'', ``X'' flags in H59) are also given.
The number of systems to a given rank
is shown by small circles connected by a rising line and by the right axis.
The star symbols mark
ranks of triple systems detected spectroscopically, but
without any indication of multiplicity in the astrometric data;
this visualizes the need for independent application of
spectroscopy for a complete picture.
Error bars represent estimated Poissonian errors. They were computed
as square root of number of systems, below and including the given rank.
}

\figcaption[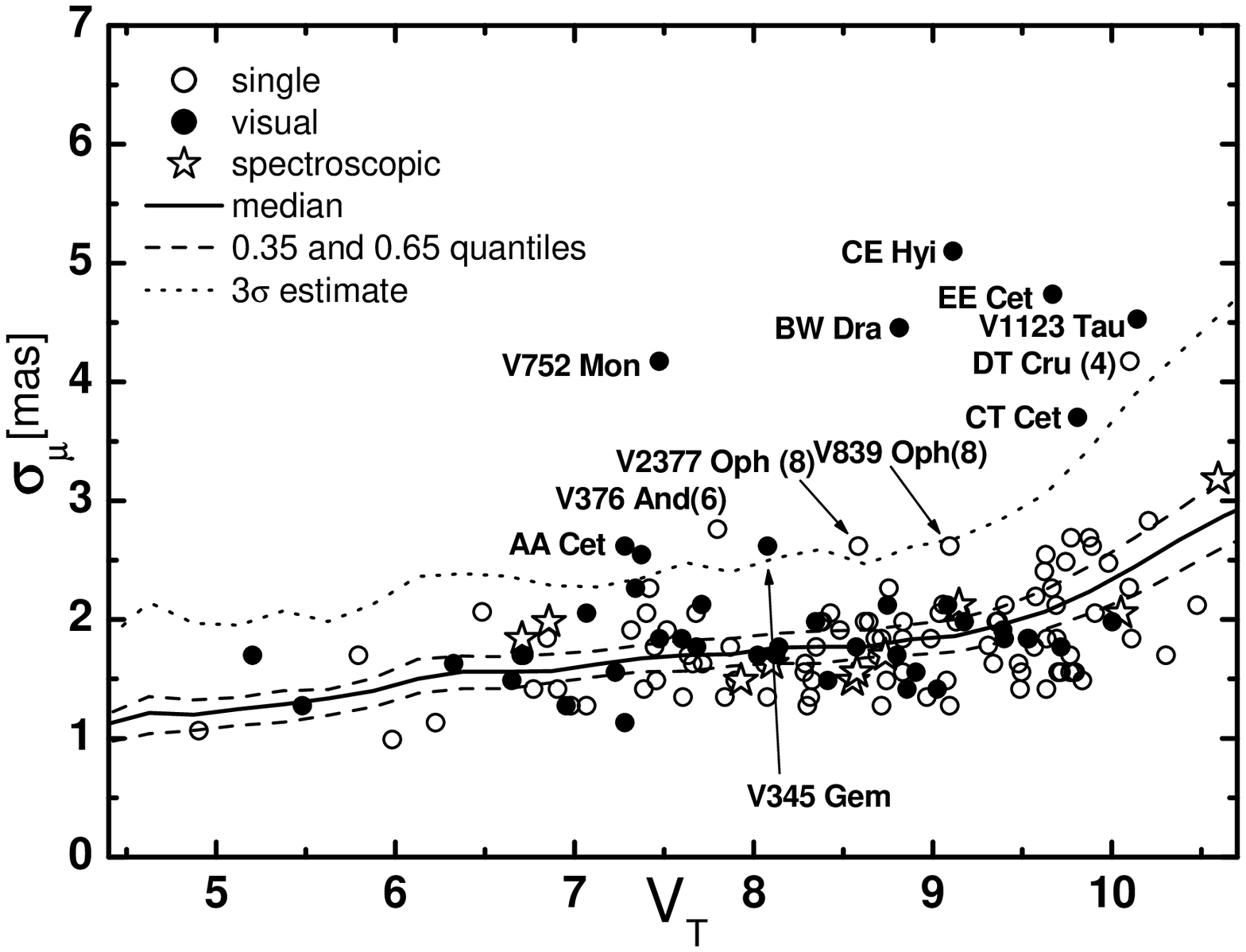]
%\figcaption[proper.eps]
{\label{err-proper} The dependence of the combined 
proper motion error, $\sigma_\mu$, on the
median $V_T$ magnitude for all contact binaries in the TYCHO-2 catalog.
This figure is analogous to Figure~\ref{errpx-V} and shares its other
symbols. All contact binaries are marked by circles with those
in known visual binaries marked by filled circles.
Numbers written in parentheses are given for
systems currently not identified as multiple systems, but with a small
number of astrometric observations available for proper motion
determination; these cases should be
treated with a considerable caution.
The star symbols mark spectroscopic binaries; it should
be noted that none would be detected by the proper-motion
error technique.
}

\figcaption[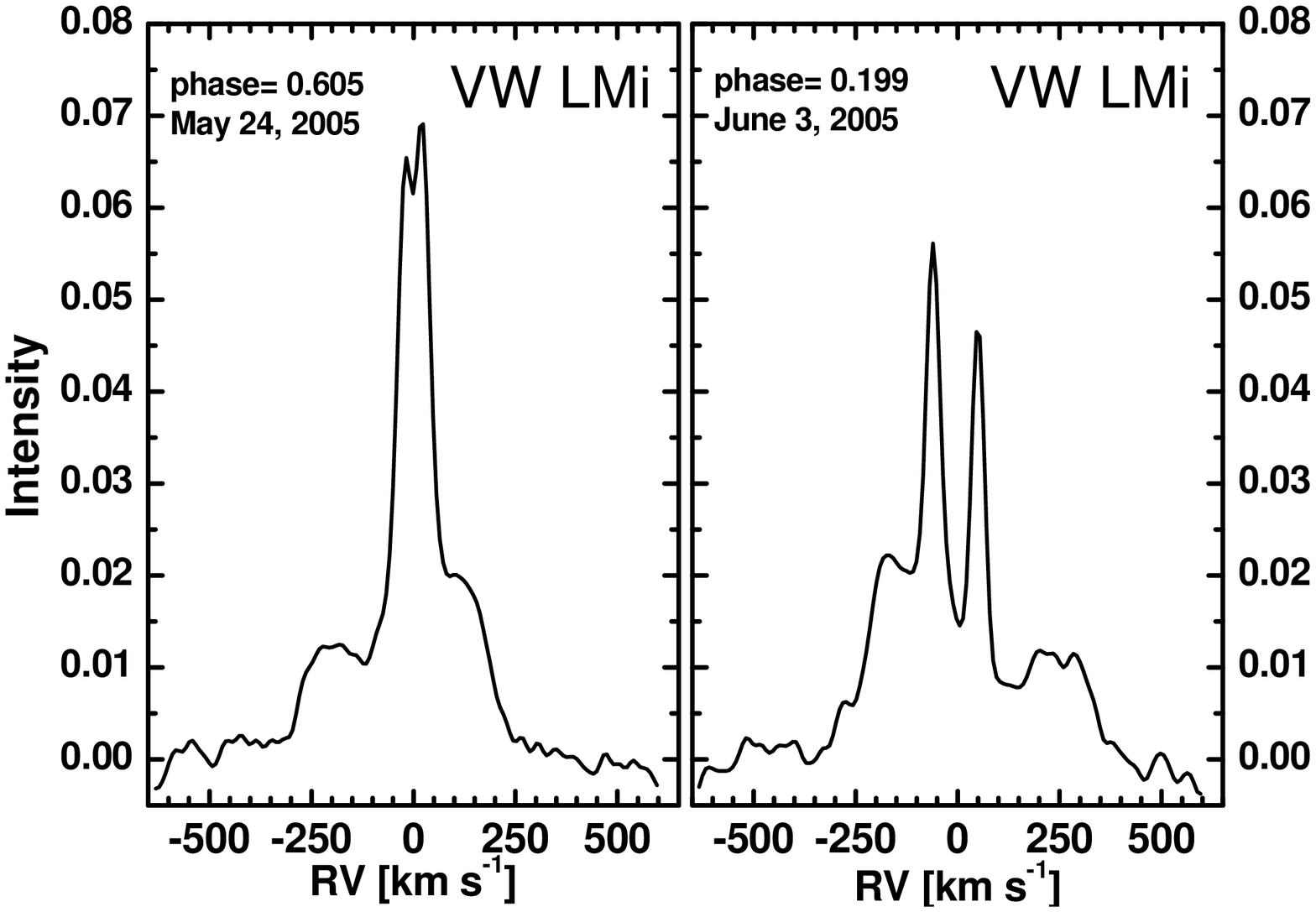]
%\figcaption[vwlmi.eps]
{\label{vwlmi} Broadening functions of the quadruple 
system VW~LMi extracted
from spectra taken on May 24, 2005 (left) and June 3, 2005 (right).
The contact binary signature is in the strongly broadened
profiles, while the profiles for the slowly rotating components
of the second, long-period, detached binary are narrow.
Splitting of the lines of the detached pair
occurs only on a few days around the periastron passage.
}

\figcaption[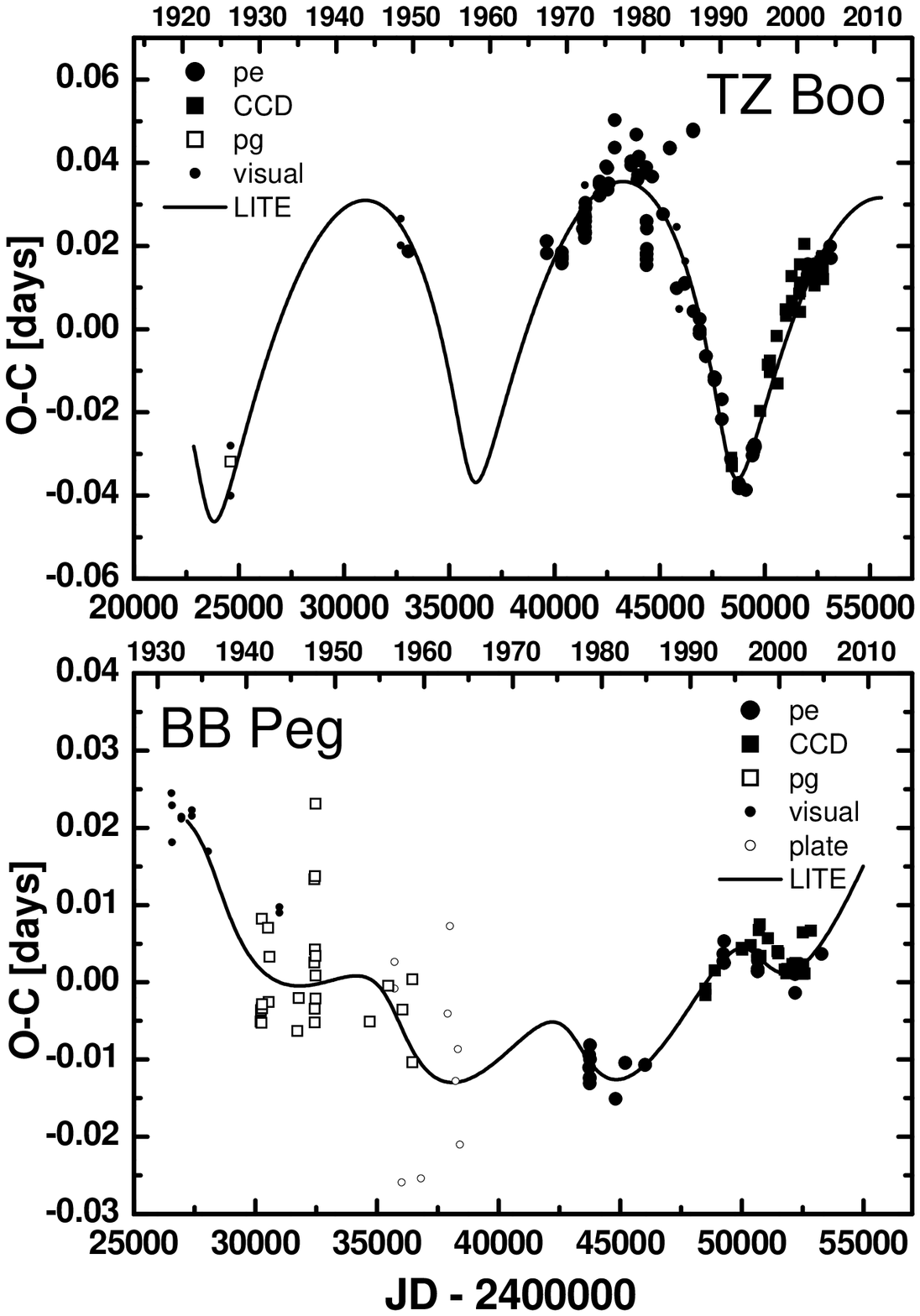]
%\figcaption[omcs.eps]
{\label{omc} Two illustrative examples of the LITE solutions:
A relatively reliably solved case of TZ~Boo and a marginal detection
for BB~Peg. The lower horizontal axis gives Julian date, the upper
horizontal axis gives the year. The vertical axis
gives (O--C) residuals from an arbitrarily selected linear ephemeris.
The values of (O--C)'s are plotted with different symbols
according to the observation type from which the instant of minimum
was derived (``CCD'' -- CCD observation, ``pe'' -- photoelectric photometry,
``pg'' -- several photographic plates, ``visual'' -- visual estimates and
``plate'' -- the time of the minimum was crudely evaluated
from the epoch of a photographic plate when the system appeared to be faint).
}

\figcaption[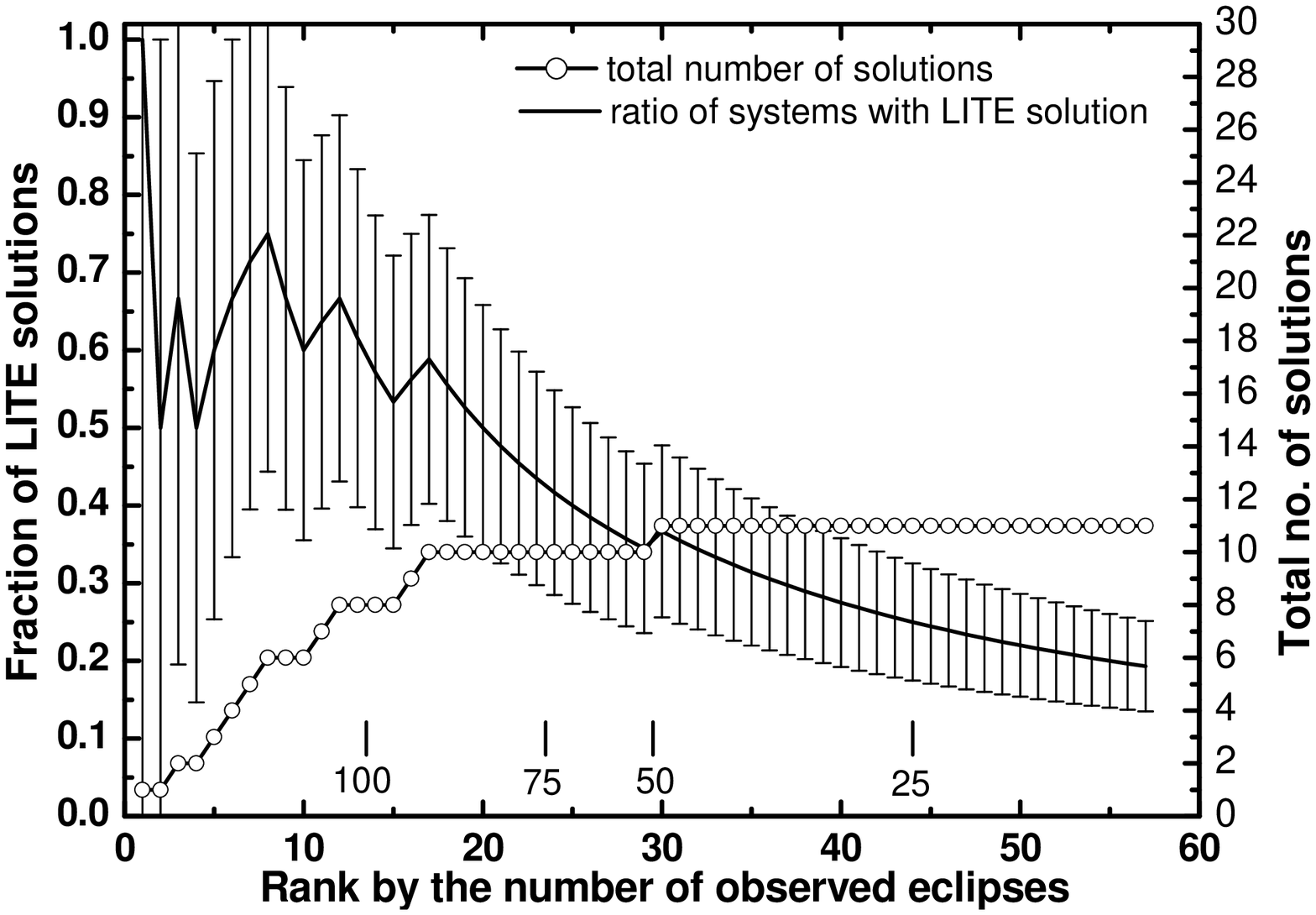]
%\figcaption[lite.eps]
{\label{litefig} The fraction of binaries having LITE solutions
with respect to the rank of the system evaluated as the (decreasing)
number of the available eclipse timings.
The error bars give the
$1-\sigma$ Poissonian uncertainty. The vertical markers
with numbers 100, 75, 50, and 25 indicate limits for the available numbers
of CCD and {\it pe\/} minima (derived from CCD and photoelectric
photometry, respectively) for the LITE solutions. The connected
circles and the right axis show the number of systems with LITE
solutions for given rank; this number increases slowly beyond the
rank of 20, which means that the detection is heavily biased by
the number of available data. The figure suggests a relatively
high incidence of multiplicity among the 20 best studied systems.
}

\figcaption[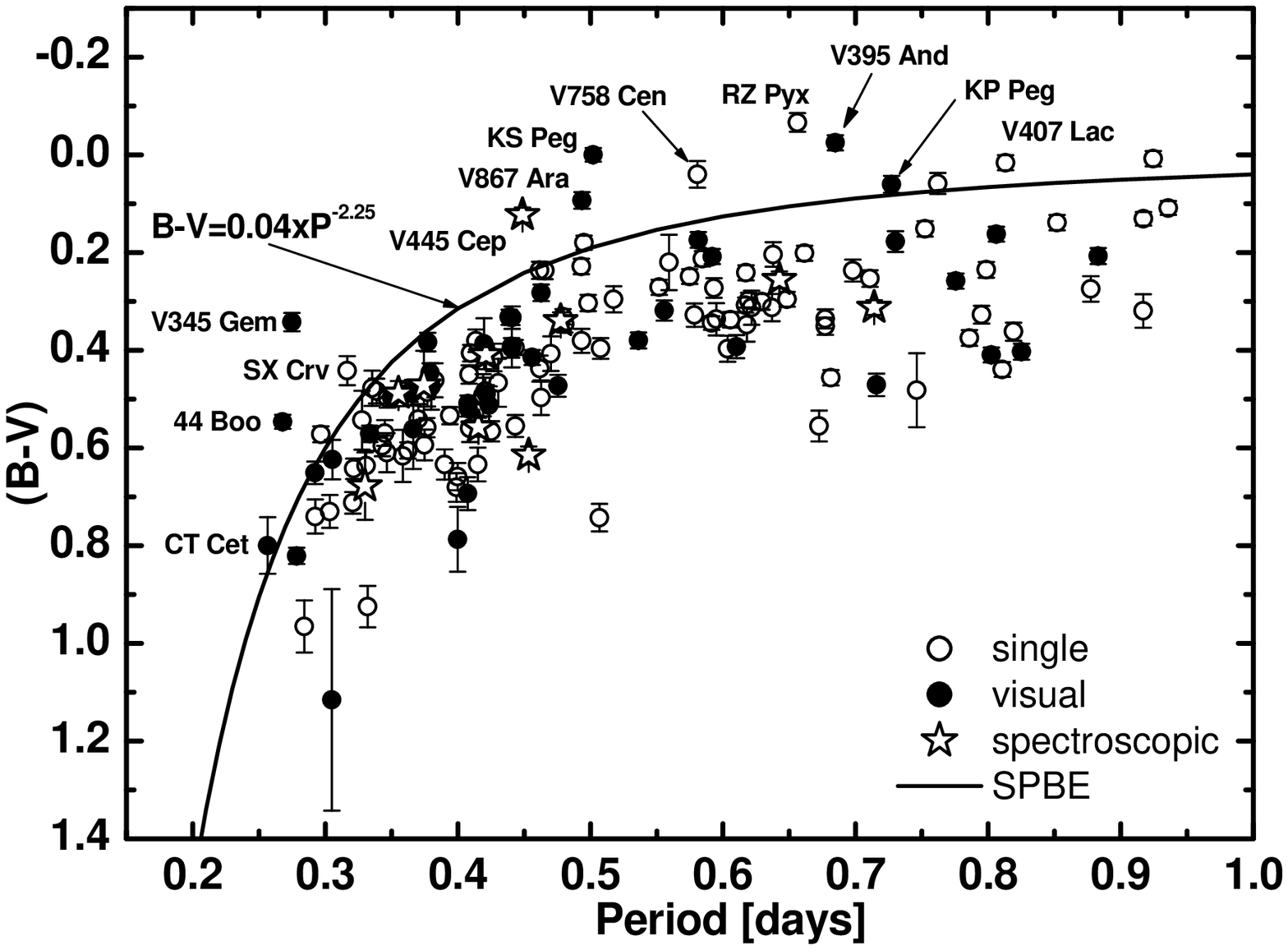]
%\figcaption[percol.eps]
{\label{spbe} The period -- color relation is shown 
for contact binaries brighter
than $V_{max} = 10$ appearing in TYCHO-2 catalogue (circles).
The filled circles show the known visual systems
while known spectroscopic triples are marked by the star symbols.
The continuous line gives a simple approximation to the Short Period Blue
Envelope (SPBE) as discussed in the text; for contact binaries,
the SPBE has a meaning similar to that of the Zero Age Main Sequence.
Systems above and to the left of the SPBE may have blue companions.
}

\figcaption[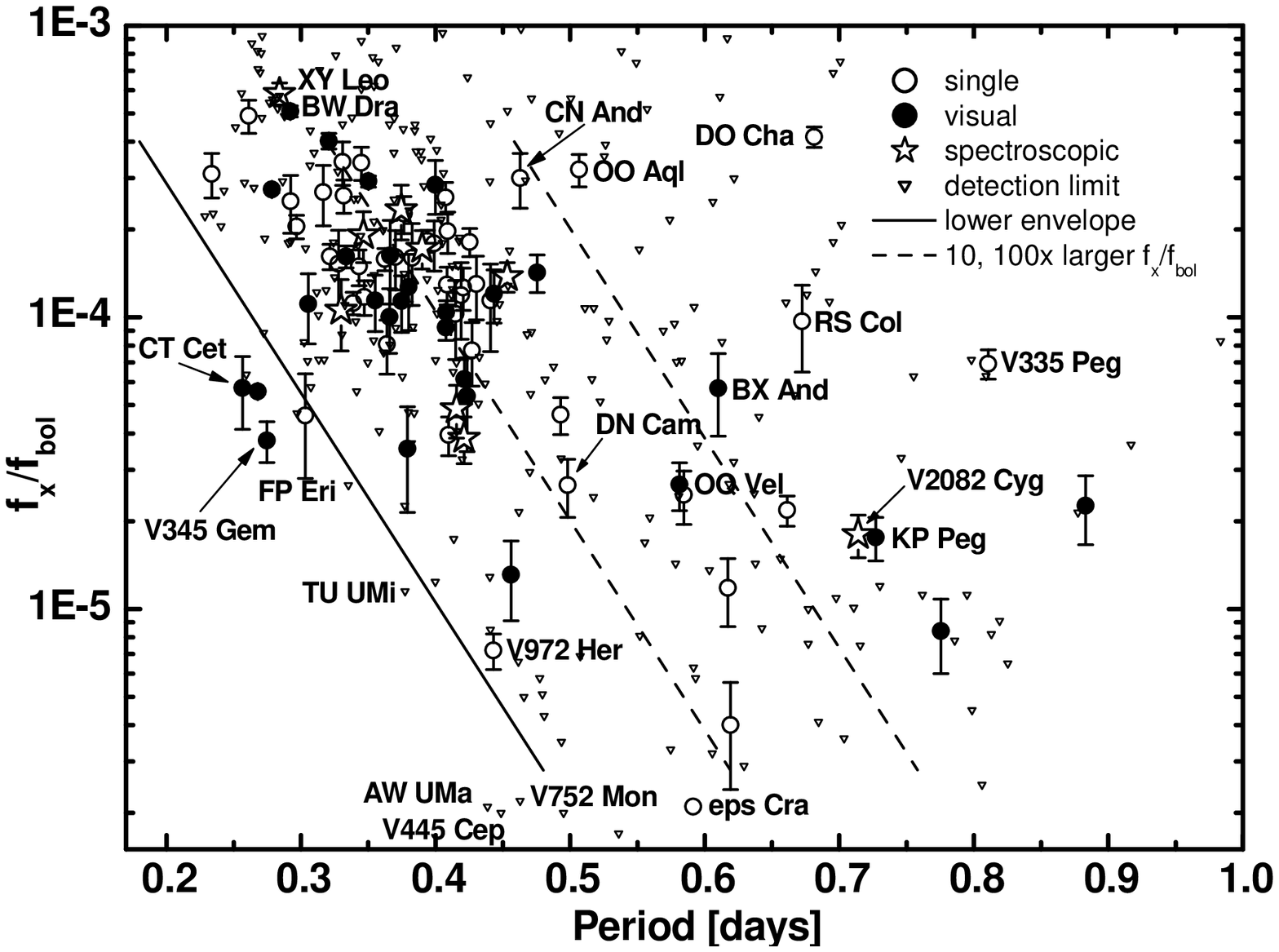]
%\figcaption[xrays.eps]
{\label{Xrays} The ratio of the apparent
X-ray and bolometric fluxes, $f_X/f_{bol}$, is expected to scale
with the orbital period with the short period systems being on the
average more X-ray active. The continuous thick line gives the
approximate locus (selected to encompass most of the systems) of
the low X-ray luminosity envelope (LXLE) in X-rays while the
broken lines give the same envelope shifted up by $10 \times$ and
$100 \times$. The $f_X/f_{bol}$ points in this figure are based on
detections (circles) and the upper limits (small triangles) for
all contact binaries in the RASS survey. Presence of companions is
expected to modify the flux ratio by enhancing it when companions
are active M-dwarfs and diminishing it when companions are early
A/F-type stars. The filled circles denote members of known visual
binaries and asterisks give spectroscopic detections (here we
added two spectroscopic detections, XY~Leo and ET~Leo, with
astrometric indication). }

\figcaption[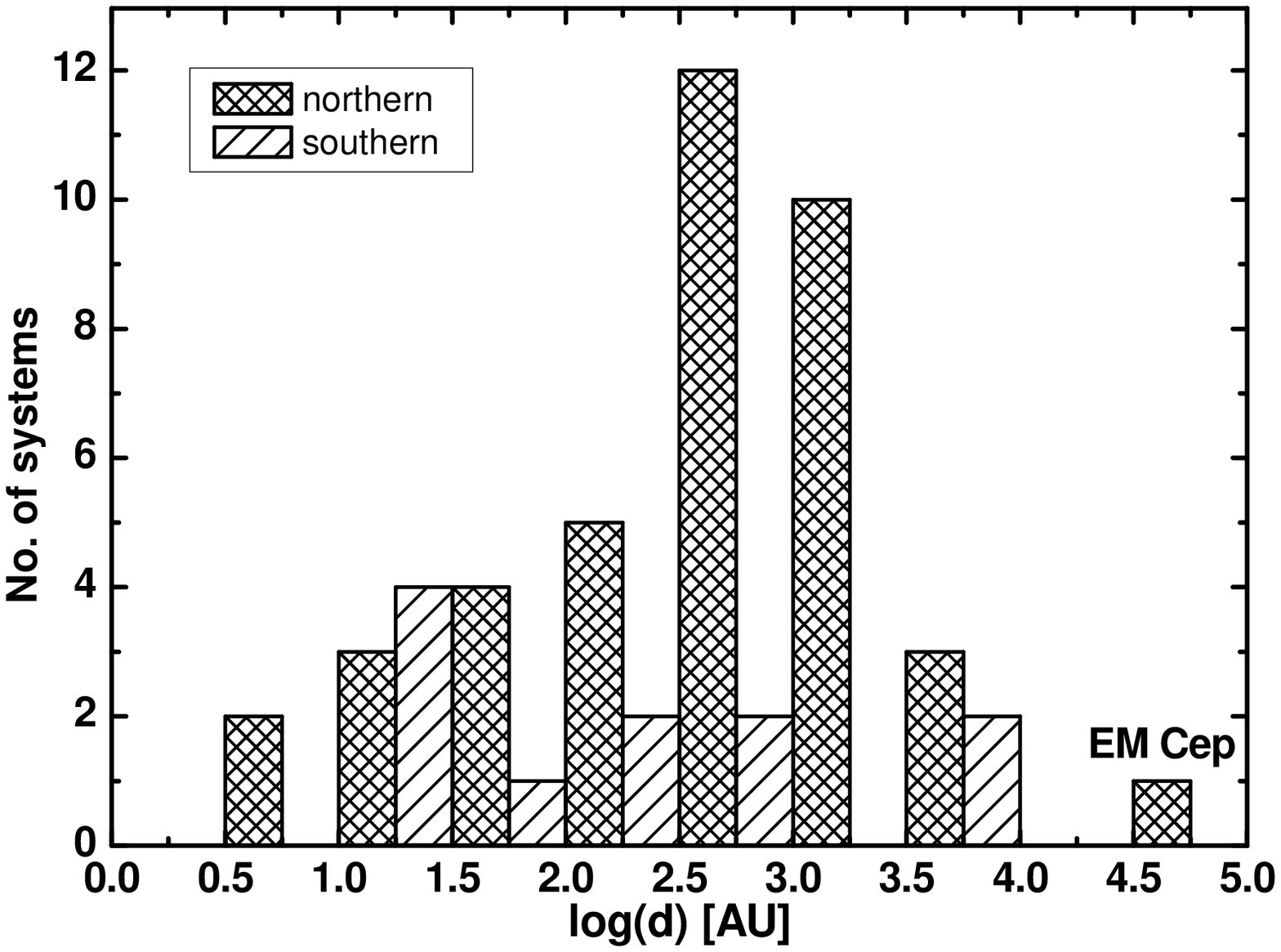]
%\figcaption[separ.eps]
{\label{separ}The projected separations in 
astronomical units are shown for all
systems with $V_{max} < 10$ with available astrometric data,
separately for both hemispheres (relative to the celestial
equator). The large disparity in numbers is directly visible.
EM~Cep (the only system in the bin with 4.5$ < \log d < 5$),
sometimes considered to be in a visual binary with a
very large separation, is probably not a physically bound system
(see the text).
}

\figcaption[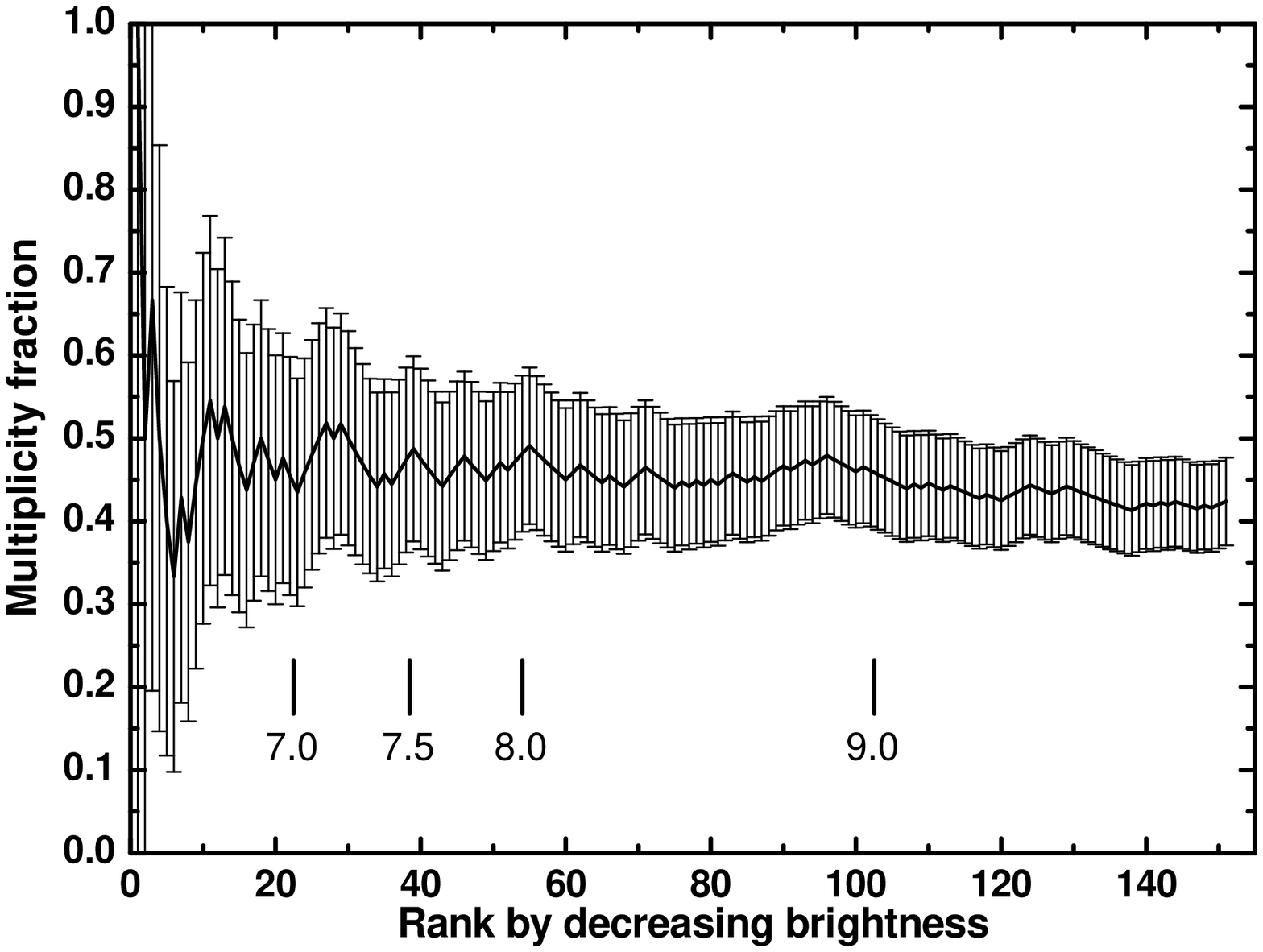]
%\figcaption[multiple.eps]
{\label{multi} The fraction of contact binaries 
having an additional component.
The systems have been ranked according to the maximum brightness.
The fraction is a cumulative one in the sense that it corresponds
to a given rank and to ranks below it. The corresponding magnitude
limits are marked by the vertical markers. Note that systems fainter
than $V_{max} \simeq 9$ appear to show a bias of a
decreasing multiplicity fraction.
}

\clearpage

\newpage
\plotone{f1.eps}
%\plotone{corr-mag.eps}

\newpage
\plotone{f2.eps}
%\plotone{visual.eps}

\newpage
\plotone{f3.eps}
%\plotone{proper.eps}

\newpage
\plotone{f4.eps}
%\plotone{vwlmi.eps}

\newpage
\plotone{f5.eps}
%\plotone{omcs.eps}

\newpage
\plotone{f6.eps}
%\plotone{lite.eps}

\newpage
\plotone{f7.eps}
%\plotone{percol.eps}

\newpage
\plotone{f8.eps}
%\plotone{xrays.eps}

\newpage
\plotone{f9.eps}
%\plotone{separ.eps}

\newpage
\plotone{f10.eps}
%\plotone{multiple.eps}

%----------------------------------------------------------------------
\begin{deluxetable}{lcrrrrrcccccccrcrcccc}

%\small
\tabletypesize{\scriptsize}

\rotate

\tabcolsep 3pt

%\tablecolumns{17}

\tablewidth{0pt}
%\tablewidth{320pt}
\tablenum{1}

\tablecaption{Multiplicity of contact binary stars: Summary data \label{summary}}

\tablehead{
\colhead{Name}            & %0
\colhead{Note}            & %1
\colhead{$V_{max}$}       & %2
\colhead{WDS}             & %3
\colhead{$\theta$}        & %4
\colhead{$\rho$}          & %5
\colhead{$d$}             & %5b
\colhead{$\Delta V$}      & %6
\colhead{HIP}             & %7
\colhead{$\sigma_\pi$}    & %9
\colhead{$\sigma_\mu$}    & %10
\colhead{$\beta$}         & %11
\colhead{$\Delta$RV}      & %12
\colhead{$P_3$}           & %13
\colhead{$a \sin i$}      & %14
\colhead{$\epsilon_{BV}$} & %15
\colhead{$\log X$}        & %16
\colhead{Moon}            & %17
\colhead{FLAGS}           & %18
\colhead{Weight}        & %19
\colhead{Triple} \\         %20
                 &          %0
                 &          %1
                 &          %2
                 &          %3
 \colhead{[deg]} &          %4
 \colhead{["]}   &          %5
\colhead{[AU]}   &          %5b
                 &          %6
                 &          %7
                 &          %9
                 &          %10
                 &          %11
                 &          %12
 \colhead{[y]}   &          %13
 \colhead{[AU]}  &          %14
                 &          %15
                 &          %16
                 &          %17
 \colhead{\tt 1234 5678} &  %18
                 &          %19
                 \\}        %20
\startdata
AB     And& *&    9.50& CFHT98     &   306& 18.90&      2266  &  7.3& {\tt -S} & 1.48&    2.69&     &     &   59& 3.8&  0.45& -3.58& -& {\tt NPSN NDNN}&    2& Y\\
BX     And&  &    8.90& 02091+4048 &    60& 19.70& {\it 4195 }& 1.37& {\tt CC} & 2.93&    1.41&     &     &   62& 2.2&  0.27& -4.24& -& {\tt P-SN -SNN}&  1.5& Y\\
CN     And&  &    9.70&            &      &      &            &     & {\tt --} &     &    1.56&     &     &     &    &  0.27& -3.52& -& {\tt N--N NNNN}&    0&  \\
GZ     And& *&    9.92& 02122+4440 &    33& 2.13 & {\it  353 }&  2.4& {\tt -S} &     &    2.76& 0.02&     &     &    &  0.19&      & -& {\tt DDSN D-NN}&  3.5& Y\\
V376   And&  &    7.63&            &      &      &            &     & {\tt --} & 0.89&    2.76&     &     &     &    &  0.17&      & -& {\tt N-NS N-NN}&  0.5&  \\
V395   And&  &    7.55& 23445+4623 &   103& 5.64 & {\it 2541 }& 2.50& {\tt CA} & 1.12&    1.77&     &     &     &    & -0.12&      & -& {\tt P-NN --SN}&    1& Y\\
S      Ant&  &    6.40&            &      &      &            &     & {\tt --} & 0.71&    2.06&     &     &     &    &  0.19&      & -& {\tt N-NN --NN}&    0&  \\
HV     Aqr&  &    9.71&            &      &      &            &     & {\tt --} &     &    2.05& 0.02&     &     &    &  0.11& -3.63& -& {\tt N--N DNNN}&    1& Y\\
OO     Aql&  &    9.20&            &      &      &            &     & {\tt --} &     &    2.19&     &     &     &    &  0.56& -3.49& -& {\tt N--N -NNL}&  0.5&  \\
V1464  Aql&  &    8.60&            &      &      &            &     & {\tt --} & 1.26&    1.84&     &     &     &    &  0.15&      & -& {\tt N-NN --NN}&    0&  \\
V535   Ara&  &    7.17&            &      &      &            &     & {\tt --} & 0.90&    2.05&     &     &     &    &  0.19&      & -& {\tt N-NN --NN}&    0&  \\
V851   Ara&  &    6.86&            &      &      &            &     & {\tt --} & 0.82&    1.41&     &     &     &    &  0.12& -4.93& -& {\tt N-NN --NN}&    0&  \\
V867   Ara&  &    7.39& 17424--4618&   316& 0.36 & {\it  55  }& 1.14& {\tt CA} & 2.62&    1.84&     &     &     &    & -0.10&      & -& {\tt D-NN --SN}&  1.5& Y\\
V870   Ara&  &    8.80&            &      &      &            &     & {\tt --} & 1.34&    2.05&     &     &     &    &  0.34&      & -& {\tt N-NN --NN}&    0&  \\
V402   Aur&  &    8.84&            &      &      &            &     & {\tt --} & 1.31&    1.35&     &     &     &    &  0.27&      & -& {\tt N-NN N-NN}&    0&  \\
V410   Aur& *&    9.94& 05012+3430 &   229& 1.70 & {\it  345 }& 1.04& {\tt CA} & 5.39&    8.41& 0.26&    9&     &    &  0.18& -3.79& -& {\tt D-NS D-NN}&  2.5& Y\\
44     Boo& *&    4.71& 15038+4739 &    49& 1.70 &       22   & 0.78& {\tt CB} & 1.03&    1.70& 1.74&  -17&     &    & -0.23& -4.25& -& {\tt D-NN DNSN}&  2.5& Y\\
CK     Boo& *&    8.99& CFHT98     &   196& 0.12 & {\it  15  }&  2.8& {\tt --} & 1.34&    2.12& 0.01&     &   21& 5.0&  0.08& -3.94& -& {\tt NDNN DSNN}&  2.5& Y\\
DW     Boo&  &    7.38&            &      &      &            &     & {\tt --} & 1.11&    2.26&     &     &     &    & -0.04&      & -& {\tt N-NN --SN}&  0.5&  \\
EF     Boo&  &    9.27&            &      &      &            &     & {\tt --} & 1.06&    1.63&     &     &     &    &  0.20& -3.90& -& {\tt N-NN N-NN}&    0&  \\
FI     Boo&  &    9.43&            &      &      &            &     & {\tt X-} & 2.10&    1.77& 0.01&     &     &    &  0.30& -3.76& -& {\tt N-DN D-NN}&    2& Y\\
AO     Cam& *&    9.50&            &      &      &            &     & {\tt --} &     &    3.18& 0.01&     &     &    &  0.19& -3.98& -& {\tt N--N DNNN}&    1& Y\\
DN     Cam&  &    8.15&            &      &      &            &     & {\tt -S} & 0.89&    1.98&     &     &     &    &  0.11& -4.58& -& {\tt N-SN N-NL}&    1& Y\\
FN     Cam&  &    8.50&            &      &      &            &     & {\tt --} & 0.95&    1.70&     &     &     &    &  0.25&      & -& {\tt N-NN N-NN}&    0&  \\
BO     CVn&  &    9.48&            &      &      &            &     & {\tt --} &     &    1.56&     &     &     &    &  0.12&      & -& {\tt N--N -NNN}&    0&  \\
IS     CMa&  &    6.84&            &      &      &            &     & {\tt --} & 0.71&    1.27&     &     &     &    &  0.19&      & -& {\tt N-NN --NN}&    0&  \\
BH     CMi&  &    9.30&            &      &      &            &     & {\tt --} &     &    1.70&     &     &     &    &  0.07&      & -& {\tt N--N -NNN}&    0&  \\
V776   Cas&  &    8.82& 01534+7003 &   137& 5.38 & {\it  885 }& 2.00& {\tt CA} & 1.56&    2.12&     &   -2&     &    &  0.14&      & -& {\tt D-NN N-NN}&    1& Y\\
RR     Cen&  &    7.27&            &      &      &            &     & {\tt --} & 0.85&    1.91&     &     &     &    &  0.21&      & -& {\tt N-NN --NN}&    0&  \\
V752   Cen&  &    9.10&            &      &      &            &     & {\tt -S} & 1.47&    1.98&     &     &     &    &  0.17& -3.79& -& {\tt N-SN --NN}&  0.5&  \\
V757   Cen&  &    8.30&            &      &      &            &     & {\tt --} & 1.10&    1.98&     &     &     &    &  0.15& -3.83& -& {\tt N-NN --NN}&    0&  \\
V758   Cen&  &    9.48&            &      &      &            &     & {\tt --} & 1.55&    2.12&     &     &     &    & -0.10&      & -& {\tt N-NN --SN}&  0.5&  \\
V759   Cen&  &    7.40&            &      &      &            &     & {\tt --} & 0.93&    1.35&     &     &     &    &  0.21& -3.74& -& {\tt N-NN -NNN}&    0&  \\
V839   Cen&  &    9.51&            &      &      &            &     & {\tt --} & 1.66&    1.84&     &     &     &    &  0.16& -3.47& -& {\tt N-NN --NN}&    0&  \\
VW     Cep& *&    7.23& 20374+7536 &   291& 0.50 &       14   & 2.80& {\tt CC} & 0.97&    1.84& 0.06&     &   28& 2.3&  0.11& -3.56& -& {\tt D-SN -DNN}&  2.5& Y\\
EM     Cep& *&    7.02& 21538+6237 &   145& 62.60& {\it 49722}& 0.84& {\tt -S} & 0.55&    2.05&     &     &     &    &  0.10&      & -& {\tt N-SN -NNN}&  0.5&  \\
V445   Cep&  &    6.81&            &      &      &            &     & {\tt --} & 0.50&    1.98& 0.06&     &     &    & -0.12&      & -& {\tt N-NN D-SE}&    2& Y\\
AA     Cet&  &    6.56& 01590--2255&   303& 8.56 & {\it  740 }& 0.36& {\tt CB} & 2.36&    2.62& 0.59&     &     &    &  0.22&      & -& {\tt D-NS --NN}&  1.5& Y\\
CL     Cet&  &    9.77&            &      &      &            &     & {\tt X-} & 3.63&    2.05&     &     &     &    &  0.20&      & -& {\tt N-DN --NN}&    1& Y\\
CT     Cet&  &    9.41& 01098--6237&    30& 3.58 & {\it  308 }& 0.79& {\tt CA} & 4.01&    3.70&     &     &     &    & -0.05& -4.24& -& {\tt D-NS --NN}&  1.5& Y\\
DY     Cet&  &    9.42&            &      &      &            &     & {\tt -S} & 2.17&    1.84&     &     &     &    &  0.14& -3.94& -& {\tt N-SN --NN}&  0.5&  \\
EE     Cet& *&    8.64& 02499+0856 &   194& 5.66 & {\it  902 }& 0.07& {\tt CA} & 4.89&    4.74&     &     &     &    &  0.12& -3.90& -& {\tt D-NS N-NN}&  1.5& Y\\
DO     Cha&  &    7.61&            &      &      &            &     & {\tt --} & 0.57&    1.63&     &     &     &    &  0.36& -3.38& -& {\tt N-NN --NL}&  0.5&  \\
RS     Col&  &    9.54&            &      &      &            &     & {\tt --} & 1.28&    2.48&     &     &     &    &  0.46& -4.01& -& {\tt N-NN --NL}&  0.5&  \\
KR     Com&  &    7.13& 13203+1746 &     8& 0.10 &        8   &  0.6& {\tt --} & 0.87&    1.56& 0.56&    4&     &    &  0.22& -3.98& -& {\tt D-NN D-NN}&    2& Y\\
eps    CrA&  &    4.74&            &      &      &            &     & {\tt --} & 0.92&    1.06&     &     &     &    &  0.21& -5.68& -& {\tt N-NN --NN}&    0&  \\
YY     CrB&  &    8.48&            &      &      &            &     & {\tt --} & 0.85&    2.26&     &     &     &    &  0.20& -3.64& -& {\tt N-NN NNNN}&    0&  \\
SX     Crv&  &    8.99&            &      &      &            &     & {\tt --} & 1.21&    1.98&     &     &     &    & -0.09& -3.57& -& {\tt NNNN NNSN}&  0.5&  \\
DT     Cru&  &    9.91&            &      &      &            &     & {\tt -S} & 1.70&    4.17&     &     &     &    &  0.27&      & -& {\tt N-SS --NN}&    1& Y\\
V1073  Cyg&  &    8.23&            &      &      &            &     & {\tt --} & 0.95&    2.05&     &     &     &    &  0.31&      & -& {\tt NNNN NNNN}&    0&  \\
V2082  Cyg&  &    6.63&            &      &      &            &     & {\tt --} & 0.56&    1.84& 0.03&     &     &    &  0.23& -4.74& -& {\tt N-NN D-NL}&  1.5& Y\\
V2150  Cyg&  &    8.00& 21182+3035 &   210& 3.68 & {\it  667 }& 3.35& {\tt CA} & 1.59&    1.77&     &     &     &    &  0.08&      & -& {\tt D-NN N-NN}&    1& Y\\
LS     Del&  &    8.61&            &      &      &            &     & {\tt --} &     &    1.27&     &     &     &    &  0.10& -4.09& -& {\tt N--N NNNN}&    0&  \\
AP     Dor&  &    9.27&            &      &      &            &     & {\tt --} & 0.92&    2.12&     &     &     &    &  0.14& -4.11& -& {\tt N-NN --NN}&    0&  \\
BV     Dra& *&    7.88& 15118+6151 &   353& 16.10&      1083  & 0.69& {\tt CA} & 2.56&    1.70&     &     &     &    &  0.07& -3.53& -& {\tt D-NN -NNN}&    1& Y\\
BW     Dra& *&    8.61& 15118+6151 &   353& 16.10& {\it 1089 }& 0.69& {\tt CA} & 5.18&    4.46&     &     &     &    &  0.01& -3.29& -& {\tt D-NS -NNN}&  1.5& Y\\
GM     Dra&  &    8.66&            &      &      &            &     & {\tt --} & 0.88&    1.98&     &     &     &    &  0.03& -3.95& -& {\tt N-NN N-NN}&    0&  \\
YY     Eri&  &    8.10&            &      &      &            &     & {\tt --} & 1.20&    1.91&     &     &     &    &  0.13& -3.79& -& {\tt NNNN -NNN}&    0&  \\
BV     Eri&  &    8.12&            &      &      &            &     & {\tt --} & 1.11&    1.77&     &     &     &    &  0.21&      & -& {\tt NNNN --NN}&    0&  \\
FP     Eri&  &    9.59&            &      &      &            &     & {\tt --} & 1.28&    1.70&     &     &     &    &  0.01&      & -& {\tt N-NN --NE}&  0.5&  \\
FX     Eri&  &    9.56&            &      &      &            &     & {\tt CA} & 1.06&    1.56&     &     &     &    &  0.10& -3.60& -& {\tt N-NN --NN}&    0&  \\
V345   Gem& *&    7.70& 07385+3343 &   352& 3.19 & {\it  233 }& 1.40& {\tt CA} & 1.77&    2.62&     &     &     &    & -0.39& -4.42& -& {\tt D-NN --SN}&  1.5& Y\\
AK     Her& *&    8.29& 17140+1621 &   323& 4.20 & {\it  425 }&    4& {\tt XS} & 2.77&    1.77&     &     &   58& 1.7&  0.21& -4.21& -& {\tt DDDN NDNN}&    4& Y\\
V842   Her&  &    9.85&            &      &      &            &     & {\tt --} &     &    2.83&     &     &     &    &  0.21&      & -& {\tt N--N N-NN}&    0&  \\
V899   Her&  &    7.82&            &      &      &            &     & {\tt --} & 0.77&    1.49& 1.50& var &     &    &  0.13& -4.41& -& {\tt N-NN D-NN}&    1& Y\\
V918   Her&  &    7.31&            &      &      &            &     & {\tt --} & 0.93&    1.41&     &     &     &    &  0.11&      & -& {\tt N-NN N-NN}&    0&  \\
V921   Her&  &    9.33&            &      &      &            &     & {\tt --} & 0.97&    1.84&     &     &     &    &  0.22&      & -& {\tt N-NN N-NN}&    0&  \\
V972   Her&  &    6.62&            &      &      &            &     & {\tt --} & 0.61&    1.70&     &     &     &    &  0.14& -5.14& -& {\tt N-NN N-NN}&    0&  \\
V1003  Her&  &    9.69&            &      &      &            &     & {\tt --} & 1.57&    1.49&     &     &     &    &  0.18&      & -& {\tt N-NN --NN}&    0&  \\
TU     Hor&  &    5.90&            &      &      &            &     & {\tt --} & 0.56&    0.99&     &     &     &    &  0.06&      & -& {\tt N-NN --NN}&    0&  \\
WY     Hor&  &    9.39&            &      &      &            &     & {\tt --} & 1.00&    1.56&     &     &     &    &  0.36& -3.75& -& {\tt N-NN --NN}&    0&  \\
FG     Hya&  &    9.90& CFHT98     &   145& 16.50& {\it 2602 }&    4& {\tt --} & 1.90&    1.84&     &     &     &    &  0.05& -3.82& -& {\tt NPNN NNNN}&  0.5&  \\
PP     Hya&  &    6.78&            &      &      &            &     & {\tt --} & 0.93&    1.84&     &     &     &    &  0.08&      & -& {\tt N-NN --NN}&    0&  \\
CE     Hyi&  &    8.36& 01389--5835&     9& 2.00 & {\it  420 }& 0.37& {\tt CA} & 2.01&    5.10&     &     &     &    &  0.08&      & -& {\tt D-NS --NN}&  1.5& Y\\
CN     Hyi&  &    6.56& 02456--7114&   132& 0.51 &       30   & 3.76& {\tt CA} & 0.65&    1.70&     &     &     &    &  0.18& -4.88& -& {\tt D-NN --NN}&    1& Y\\
CP     Hyi&  &    7.80&            &      &      &            &     & {\tt --} & 0.60&    1.49&     &     &     &    &  0.12&      & -& {\tt N-NN --NN}&    0&  \\
SW     Lac& *&    8.51& CFHT98     &    85& 1.70 &       138  & 2.54& {\tt -S} & 1.26&    1.84& 0.01& sim &   28& 2.6&  0.20& -3.40& -& {\tt NDSN DDNN}&  3.5& Y\\
V407   Lac&  &    8.29&            &      &      &            &     & {\tt --} & 0.93&    1.49&     &     &     &    & -0.05&      & -& {\tt N-NN --SN}&  0.5&  \\
UZ     Leo&  &    9.58&            &      &      &            &     & {\tt -S} & 1.59&    1.56&     &     &     &    &  0.23&      & M& {\tt NNSN NNNN}&  0.5&  \\
XY     Leo& *&    9.45&            &      &      &            &     & {\tt -S} & 1.80&    1.56& 0.10& var &   20& 4.1&  0.29& -3.23& M& {\tt NNSN DDNN}&  2.5& Y\\
AM     Leo& *&    9.25& 11022+0954 &   270& 11.50&       883  & 1.68& {\tt CA} & 3.64&    1.91&     &     &     &    &  0.10& -4.00& M& {\tt D-NN -NNN}&    1& Y\\
AP     Leo&  &    9.32&            &      &      &            &     & {\tt -S} & 1.70&    2.40&     &     &     &    &  0.20& -3.89& M& {\tt NNSN NNNN}&  0.5&  \\
ET     Leo&  &    9.44&            &      &      &            &     & {\tt -S} & 1.44&    1.41& 0.02&     &     &    &  0.18& -3.72& -& {\tt N-SN D-NN}&  1.5& Y\\
EX     Leo&  &    8.13&            &      &      &            &     & {\tt --} & 1.11&    1.27&     &     &     &    &  0.15& -3.89& -& {\tt N-NN N-NN}&    0&  \\
VW     LMi& *&    7.93&            &      &      &            &     & {\tt --} & 0.90&    1.63& 0.42& var &     &    &  0.13&      & -& {\tt N-NN D-NN}&    1& Y\\
ES     Lib&  &    7.11& 15168-1302 &   174& 0.20 &       25   & ?   & {\tt -S} & 0.91&    1.13&     &     &     &    &  0.15& -4.64& M& {\tt D-SN --NN}&  1.5& Y\\
FT     Lup&  &    9.70&            &      &      &            &     & {\tt --} &     &    2.69&     &     &     &    &  0.19&      & -& {\tt N--N -NNN}&    0&  \\
UV     Lyn&  &    9.41&            &      &      &            &     & {\tt --} & 1.62&    2.55&     &     &     &    &  0.34& -3.99& -& {\tt NNNN NNNN}&    0&  \\
TY     Men&  &    8.08&            &      &      &            &     & {\tt --} & 0.63&    1.63&     &     &     &    &  0.01&      & -& {\tt N-NN -NNN}&    0&  \\
AN     Men&  &    9.24&            &      &      &            &     & {\tt --} & 0.86&    1.98&     &     &     &    &  0.21&      & -& {\tt N-NN --NN}&    0&  \\
V752   Mon& *&    6.90& 07079--0441&    26& 1.71 & {\it  213 }& 0.51& {\tt CA} & 1.40&    4.17&     &     &     &    &  0.06&      & -& {\tt D-NS --NN}&  1.5& Y\\
V753   Mon&  &    8.21&            &      &      &            &     & {\tt --} & 1.04&    1.35&     &     &     &    &  0.24&      & -& {\tt N-NN N-NN}&    0&  \\
V2     Mon&  &    9.90&            &      &      &            &     & {\tt --} &     &    2.48&     &     &     &    &  0.21& -3.98& -& {\tt N--N --NN}&    0&  \\
V3     Mon&  &    8.90&            &      &      &            &     & {\tt --} &     &    1.63&     &     &     &    &  0.09&      & -& {\tt N--N --NN}&    0&  \\
DE     Oct&  &    9.09& 20194--7608&   128& 22.90& {\it 5423 }& 2.85& {\tt --} & 1.03&    1.98&     &     &     &    &  0.17&      & -& {\tt P-NN --NN}&  0.5&  \\
V502   Oph& *&    8.34&            &      &      &            &     & {\tt --} & 1.17&    1.56& 0.01& sim &     &    &  0.38& -3.86& -& {\tt NNNN DNNN}&    1& Y\\
V566   Oph&  &    7.46&            &      &      &            &     & {\tt --} & 1.11&    1.70&     &     &   23& 1.2&  0.11& -4.40& -& {\tt NNNN NDNN}&    1& Y\\
V839   Oph&  &    8.80& CFHT98     &     8& 8.20 &      1014  &  6.1& {\tt --} & 1.44&    2.62&     &     &     &    &  0.26& -3.70& -& {\tt NDNS NNNN}&  1.5& Y\\
V1010  Oph&  &    6.04&            &      &      &            &     & {\tt --} & 0.83&    1.13&     &     &     &    &  0.10&  4.66& -& {\tt N-NN N-NN}&    0&  \\
V2377  Oph&  &    8.45&            &      &      &            &     & {\tt --} & 1.22&    2.62&     &     &     &    &  0.29& -3.74& -& {\tt N-NS N-NN}&  0.5&  \\
V2388  Oph& *&    6.14& 17542+1108 &      & 0.09 &        6   &  0.2& {\tt G-} & 0.81&    1.63& 0.20&   -5&     &    &  0.34&      & -& {\tt D-DN D-NN}&    3& Y\\
ER     Ori&  &    9.28& 05112--0833&   354& 0.19 & {\it  31  }& 1.99& {\tt G-} & 2.16&    1.84&     &     &   52& 5.7&  0.24& -4.27& -& {\tt DDDN -SNN}&  3.5& Y\\
V1387  Ori&  &    8.71& 06094+1711 &    12& 3.19 & {\it 1003 }& 3.48& {\tt CC} & 2.23&    1.41&     &     &     &    &  0.10&      & -& {\tt D-SN --NN}&  1.5& Y\\
MW     Pav&  &    8.63&            &      &      &            &     & {\tt --} & 1.08&    1.56&     &     &     &    &  0.26&      & -& {\tt N-NN --NN}&    0&  \\
V386   Pav&  &    8.28&            &      &      &            &     & {\tt --} & 1.01&    1.98&     &     &     &    &  0.12&      & -& {\tt N-NN --NN}&    0&  \\
U      Peg&  &    9.23& CFHT98     &   275& 4.05 &       564  &  5.2& {\tt --} & 1.43&    1.56&     &     &     &    &  0.23& -3.94& -& {\tt NDNN -NNN}&    1& Y\\
KP     Peg&  &    7.05& 21267+1341 &   217& 3.46 & {\it  637 }& 1.77& {\tt CA} & 1.67&    2.55& 0.22&     &     &    & -0.02& -4.76& -& {\tt D-NS D-SN}&    3& Y\\
KS     Peg&  &    5.45& 23379+1824 &   246& 27.70&      2029  & 5.20& {\tt --} & 0.75&    1.27&     &     &     &    & -0.19&      & -& {\tt P-NN N-SN}&    1& Y\\
V335   Peg&  &    7.23&            &      &      &            &     & {\tt --} & 0.86&    1.91&     &     &     &    &  0.38& -4.16& -& {\tt N-NN N-NL}&  0.5&  \\
V351   Peg&  &    7.92&            &      &      &            &     & {\tt --} & 0.92&    1.35&     &     &     &    &  0.14& -4.60& -& {\tt N-NN N-NN}&    0&  \\
V353   Peg&  &    7.41&            &      &      &            &     & {\tt G-} & 0.89&    1.77&     &     &     &    &  0.08&      & -& {\tt N-DN --NN}&    1& Y\\
V357   Peg&  &    8.91&            &      &      &            &     & {\tt --} & 1.21&    1.27&     &     &     &    &  0.19&      & -& {\tt N-NN --NN}&    0&  \\
V2     Peg&  &    9.50&            &      &      &            &     & {\tt --} &     &    1.41&     &     &     &    &  0.20&      & -& {\tt N--N --NN}&    0&  \\
IW     Per&  &    5.76&            &      &      &            &     & {\tt --} & 0.81&    1.70&     &     &     &    &  0.08&      & -& {\tt N-NN N-NN}&    0&  \\
V579   Per&  &    7.78&            &      &      &            &     & {\tt --} & 0.89&    1.77&     &     &     &    &  0.01&      & -& {\tt N-NN --NN}&    0&  \\
V592   Per&  &    8.19& 04445+3953 &   194& 0.21 & {\it  39  }& 0.85& {\tt CA} & 1.55&    1.49& 0.60&    1&     &    &  0.39&      & -& {\tt D-NN D-NN}&    2& Y\\
AE     Phe&  &    7.56&            &      &      &            &     & {\tt --} & 0.81&    1.35&     &     &     &    &  0.21& -3.80& -& {\tt N-NN -NNN}&    0&  \\
BM     Phe&  &    9.16&            &      &      &            &     & {\tt --} & 1.17&    1.78&     &     &     &    &  0.09&      & -& {\tt N-NN --NN}&    0&  \\
TV     Pic&  &    7.37&            &      &      &            &     & {\tt --} & 0.63&    1.49&     &     &     &    &  0.08&      & -& {\tt N-NN --NN}&    0&  \\
AQ     Psc&  &    8.60& 01211+0736 &    64& 26.20&      3263  &  2.8& {\tt -S} & 1.29&    2.12&     &     &     &    &  0.26& -3.85& M& {\tt PPSN N-NN}&  1.5& Y\\
TY     Pup&  &    8.40& CFHT98     &     2& 8.40 &      4264  &     & {\tt --} & 1.01&    1.98&     &     &     &    &  0.30&      & -& {\tt NPNN --NN}&  0.5&  \\
RZ     Pyx&  &    8.94&            &      &      &            &     & {\tt --} & 1.06&    2.05&     &     &     &    & -0.17&      & -& {\tt N-NN --SN}&  0.5&  \\
V701   Sco&  &    8.63&            &      &      &            &     & {\tt --} & 1.31&    1.84&     &     &     &    & -0.02&      & -& {\tt N-NN --NN}&    0&  \\
V1084  Sco&  &    8.92&            &      &      &            &     & {\tt --} & 1.32&    2.13&     &     &     &    &  0.14& -4.34& -& {\tt N-NN --NN}&    0&  \\
OU     Ser&  &    8.10&            &      &      &            &     & {\tt --} & 0.95&    1.56&     &     &     &    & -0.04& -3.69& -& {\tt N-NN N-SN}&  0.5&  \\
Y      Sex&  &    9.83& 07385+3343 &   154& 0.49 & {\it  121 }& 2.62& {\tt CB} & 3.38&    1.98&     &     &     &    &  0.10&      & -& {\tt D-NN -NNN}&    1& Y\\
VY     Sex&  &    9.01&            &      &      &            &     & {\tt --} &     &    1.49&     &     &     &    &  0.31& -3.92& -& {\tt N--N N-NN}&    0&  \\
V781   Tau&  &    8.56& CFHT98     &   163& 8.00 &       650  &    8& {\tt --} & 1.35&    1.49&     &     &     &    &  0.13& -3.47& M& {\tt NDNN -NNN}&    1& Y\\
V1123  Tau&  &    9.70& 03350+1743 &   135& 4.27 & {\it  571 }& 1.73& {\tt CA} & 5.37&    4.53&     &     &     &    &  0.47& -3.55& M& {\tt D-NS --NN}&  1.5& Y\\
V1128  Tau&  &    9.46& 03495+1255 &   196& 12.10& {\it 1310 }& 1.08& {\tt CA} & 4.77&    1.77&     &     &     &    &  0.05& -3.95& -& {\tt D-NN --NN}&    1& Y\\
AQ     Tuc&  &    9.91&            &      &      &            &     & {\tt --} & 1.24&    2.27&     &     &     &    &  0.21&      & -& {\tt N-NN --NN}&    0&  \\
DX     Tuc&  &    9.47&            &      &      &            &     & {\tt --} & 1.12&    2.26&     &     &     &    &  0.13&      & -& {\tt N-NN --NN}&    0&  \\
W      UMa&  &    7.75& 09438+5557 &    43& 6.40 &       317  & 4.45& {\tt -S} & 1.05&    1.70&     &     &     &    &  0.10& -3.79& -& {\tt D-SN SNNN}&  1.5& Y\\
AW     UMa& *&    6.83& 11301+2958 &    33& 67.20&      4442  &  2.5& {\tt --} & 0.90&    1.27&     &     &   17& 0.4&  0.08&      & -& {\tt P-NN SDNN}&  1.5& Y\\
HN     UMa&  &    9.75&            &      &      &            &     & {\tt --} & 1.39&    2.62&     &     &     &    &  0.11& -3.80& -& {\tt N-NN N-NN}&    0&  \\
HV     UMa&  &    8.52&            &      &      &            &     & {\tt --} & 1.23&    1.84&     &     &     &    &  0.17&      & -& {\tt N-NN --NN}&    0&  \\
HX     UMa&  &    8.75& 12021+4304 &    86& 0.63 & {\it  78  }& 3.31& {\tt CB} & 3.01&    1.56& 0.05&   -4&     &    &  0.09& -4.45& -& {\tt D-NN D-NN}&    2& Y\\
II     UMa&  &    8.04& 12329+5448 &     9& 0.87 & {\it  178 }& 1.64& {\tt CA} & 1.81&    1.98& 0.17&   -8&     &    &  0.34&      & -& {\tt D-NN D-NN}&    2& Y\\
TU     UMi&  &    8.73& 14557+7618 &   141& 0.22 & {\it  46  }& 0.47& {\tt CA} & 0.97&    1.70& 1.25&     &     &    &  0.02&      & -& {\tt D-NN D-NN}&    2& Y\\
TV     UMi& *&    8.62&            &      &      &            &     & {\tt --} & 0.78&    1.70& 0.90& var &     &    &  0.27& -4.31& -& {\tt N-NN D-NN}&    1& Y\\
OQ     Vel&  &    7.67& 08464--5251&   334& 0.10 &            & 0.50& {\tt --} & 0.62&    2.13&     &     &     &    &  0.18& -4.57& -& {\tt N-NN --NN}&    0&  \\
LL     Vel&  &    6.72&            &      & <0.03&       19   &     & {\tt -S} & 0.60&    1.41&     &     &     &    &  0.04& -5.40& -& {\tt D-SN --NN}&  1.5& Y\\
V353   Vel&  &    6.79&            &      &      &            &     & {\tt --} & 0.83&    1.63&     &     &     &    & -0.01&      & -& {\tt N-NN --NN}&    0&  \\
AG     Vir&  &    8.35&            &      &      &            &     & {\tt --} & 1.18&    1.49& 0.05& sim &     &    &  0.15&      & -& {\tt NNNN DNNN}&    1& Y\\
AH     Vir& *&    8.89& 12143+1149 &    16& 1.73 & {\it  174 }& 3.00& {\tt CC} & 3.11&    1.84& 0.11&     &   41& 4.2&  0.39& -3.59& -& {\tt DDSN DDNN}&  4.5& Y\\
CX     Vir&  &    9.80&            &      &      &            &     & {\tt --} &     &    2.12&     &     &     &    &  0.40&      & M& {\tt N--N --NN}&    0&  \\
FO     Vir&  &    6.50& 13298+0106 &   336& 22.80&      1926  &  6.4& {\tt -S} & 0.85&    1.49&     &     &     &    &  0.19& -5.08& -& {\tt D-SN N-NN}&  1.5& Y\\
GR     Vir&  &    7.80&            &      &      &            &     & {\tt --} & 1.18&    1.63&     &     &     &    &  0.07& -3.93& -& {\tt NNNN NNNN}&    0&  \\
HT     Vir& *&    7.06& 13461+0507 &   163& 0.56 &       36   & 0.63& {\tt CB} & 2.72&    2.26& 1.06& var &     &    &  0.21& -4.03& -& {\tt D-NN D-NN}&    2& Y\\
MW     Vir&  &    6.93&            &      &      &            &     & {\tt --} & 0.91&    1.27&     &     &     &    &  0.03& -4.33& -& {\tt N-NN --NN}&    0&  \\
NN     Vir&  &    7.60&            &      &      &            &     & {\tt --} & 1.14&    2.05&     &     &     &    &  0.14&      & -& {\tt N-NN N-NN}&    0&  \\
\hline
AH     Aur&  &   10.20& CFHT98     &    63& 3.20 &       518  & 4.20& {\tt -S} & 2.05&    2.40&     &     &     &    &  0.37& -3.71& M& {\tt NDSN NNNN}&  1.5& Y\\
TY     Boo&  &   10.81&            &      &      &            &     & {\tt --} &     &    6.15&     &     &   56& 4.1&  0.07&      & -& {\tt N--S -SNN}&    1& Y\\
TZ     Boo&  &   10.41& CFHT98     &   120& 10.60& {\it 1885 }& 6.90& {\tt --} & 1.49&    1.84&     &     &   34& 6.3& -0.05&      & -& {\tt NDNN -DSE}&    3& Y\\
TX     Cnc&  &   10.00& 08400+1900 &    25& 25.10& {\it 3765 }&  2.4& {\tt --} &     &    1.77&     &     &     &    &  0.36& -3.89& M& {\tt P--N NNNN}&  0.5&  \\
AH     Cnc& *&   13.31& 08517+1151 &   273& 26.80& {\it 23970}& 0.24& {\tt --} &     &        &     &     &     &    &      & -3.63& -& {\tt N--- --NN}&    0&  \\
V523   Cas&  &   10.62&            &      &      &            &     & {\tt --} &     &    3.61&     &     &   86& 4.8& -0.28& -3.51& -& {\tt N--N NSSN}&    1& Y\\
GW     Cep&  &   11.40&            &      &      &            &     & {\tt --} &     &    2.90&     &     &   14& 0.7& -0.05&      & -& {\tt N--N -DSN}&  1.5& Y\\
TW     Cet&  &   10.43& 01489--2053&   296& 8.79 & {\it 1588 }& 1.99& {\tt CB} & 3.50&    3.47&     &     &     &    &  0.07& -3.52& -& {\tt D-NN --NN}&    1& Y\\
RW     Com&  &   11.00&            &      &      &            &     & {\tt -S} & 2.45&    3.54&     &     &     &    &  0.06& -3.82& -& {\tt NNSN -NNN}&  0.5&  \\
CV     Cyg&  &   10.80& 19543+3803 &   140& 0.70 & {\it  787 }& 0.02& {\tt --} &     &    2.40&     &     &     &    &  0.34&      & -& {\tt D--N --NN}&    1& Y\\
DK     Cyg&  &   10.37&            &      &      &            &     & {\tt -S} & 1.78&    2.19&     &     &     &    &  0.17&      & -& {\tt NNSN NNNN}&  0.5&  \\
V401   Cyg&  &   10.64& CFHT98     &   359& 18.00& {\it 9262 }&  1.2& {\tt X-} & 8.04&    2.62& 0.03& sim &     &    &  0.17&      & -& {\tt NPDN D-NN}&  2.5& Y\\
V700   Cyg&  &   11.90& 20311+3847 &    74& 21.60& {\it 6317 }&  1.5& {\tt --} &     &        &     &     &     &    &      &      & M& {\tt P--- --NN}&  0.5&  \\
EF     Dra&  &   10.48& 18055+6945 &   240& 2.40 & {\it  807 }& 1.19& {\tt --} &     &    3.68& 0.20&    4&     &    &  0.18& -3.82& -& {\tt D--N DNNN}&    2& Y\\
UX     Eri&  &   10.50&            &      &      &            &     & {\tt -S} & 2.84&    3.18&     &     &   43& 3.9&  0.36& -3.30& -& {\tt NNSN NSNN}&    1& Y\\
QW     Gem&  &   10.18& 06508+2927 &    26& 6.38 & {\it 1567 }& 1.16& {\tt CA} & 5.42&    2.69&     &     &     &    &  0.07& -4.04& -& {\tt D-NN N-NN}&    1& Y\\
V829   Her&  &   10.10&            &      &      &            &     & {\tt --} &     &    4.03&     &     &   11& 0.9&  0.11& -3.76& -& {\tt N--S NSNN}&    1& Y\\
XZ     Leo&  &   10.18& CFHT98     &   112& 18.30& {\it 5801 }&    7& {\tt -S} & 1.71&    1.70&     &     &     &    &  0.19&      & M& {\tt NPSN NNNN}&    1& Y\\
VZ     Lib&  &   10.13&            &      &      &            &     & {\tt --} & 1.96&    2.55& 0.20& var &     &    &  0.33&      & M& {\tt NNNN D-NN}&    1& Y\\
V532   Mon&  &   12.20& 07020--0018&   179& 10.90& {\it 10035}&  0.1& {\tt --} &     &    2.48&     &     &     &    &      &      & -& {\tt P--N --NN}&  0.5&  \\
V508   Oph&  &   10.06& CFHT98     &    19& 2.40 & {\it  419 }&  3.9& {\tt -S} & 2.14&    2.19&     &     &     &    &  0.13& -3.92& -& {\tt NDSN -NNN}&  1.5& Y\\
V1363  Ori&  &   10.22&            &      &      &            &     & {\tt -S} & 2.36&    1.41&     &     &     &    &  0.30&      & -& {\tt N-SN N-NN}&  0.5&  \\
BB     Peg& *&   10.80&            &      &      &            &     & {\tt --} & 2.26&    2.69& 0.01&     &   21& 0.8&  0.09&      &  & {\tt NNNN DSNN}&  1.5& Y\\
KN     Per&  &   11.52&            &      &      &            &     & {\tt --} & 3.49&    2.48&     &     &     &    &  0.29&      & -& {\tt N-NN --NN}&    0&  \\
BQ     Phe&  &   10.32&            &      &      &            &     & {\tt -S} & 2.08&    2.33&     &     &     &    &  0.27&      & -& {\tt N-SN --NN}&  0.5&  \\
CW     Sge&  &   11.13&            &      &      &            &     & {\tt XS} & 4.14&    2.26&     &     &     &    &  0.28&      & -& {\tt N-DN --NN}&    1& Y\\
BE     Scl&  &   10.24&            &      &      &            &     & {\tt X-} & 5.11&    2.48&     &     &     &    &  0.29&      & -& {\tt N-DN --NN}&    1& Y\\
RS     Ser&  &   10.80&            &      &      &            &     & {\tt --} &     &    2.33&     &     &     &    &  0.39& -3.52& -& {\tt N--N --NL}&  0.5&  \\
RZ     Tau&  &   10.08& CFHT98     &    43& 0.80 & {\it  139 }&  3.7& {\tt -S} & 1.85&    1.84&     &     &     &    &  0.11&      & -& {\tt NDSN -NNN}&  1.5& Y\\
EQ     Tau&  &   10.50&            &      &      &            &     & {\tt --} &     &    2.05&     &     &   50& 3.9&  0.38&      & M& {\tt N--N NSNN}&  0.5&  \\
\enddata \tablecomments{
The second part of the table below the horizontal line contains binaries fainter than $V=10$.\\[1ex]
Explanation of columns: Name -- as in the contact binary in the
General Catalog of Variable Stars; Note -- an asterisk marks a
comment in Table~\ref{notes}; $V_{max}$ -- Johnson or transformed
Hipparcos $H_p$ maximum magnitude; WDS -- designation of the
system (if it is visual binary or common-proper-motion pair) in
the most recent on-line version of the catalogue
(http://ad.usno.navy.mil/wds/wds.html) or CFHT indicating a new
detection, as given in Table~\ref{cfht}; $\theta$ -- position
angle of the secondary (= fainter) component in degrees; $\rho$ --
angular separation of the components [arcsec]; $d$ -- projected
separation in astronomical units computed from angular separation
and Hipparcos parallax, if relative error of parallax larger than
0.2 or negative parallax distance (separation typed in italics)
estimated using absolute magnitude calibration of \citet{rd1997}
or estimated otherwise (see text); $\Delta V$ - visual brightness
difference of the components; HIP -- Hipparcos flags H59 and H61
(H59, type of astrometric solution: C -- linear motion, G --
acceleration term necessary, i.e., very probably an astrometric
binary, X -- stochastic solution, i.e. very probably an
astrometric binary with $P < 3$ years; H61, quality of the orbital
solution: grades ABCD (A = best) or suspected non-single flag S);
$\sigma_\pi$ -- Hipparcos parallax error [mas]; $\sigma_\mu =
\sqrt{(\sigma (\mu_\alpha) \cos \delta)^2 + \sigma^2
(\mu_\delta)}$ -- proper-motion error in the TYCHO 2 catalogue;
$\beta$ -- the ratio of apparent luminosities $l_3/(l_1+l_2)$
determined from broadening functions (see the text); $\Delta RV$
-- the difference of the third component velocity and systemic
velocity of the underlying binary [km~s$^{-1}$], ``sim'' signifies
similar velocities;
 $P_3$ -- the orbital period of the third component as determined
from LITE; $a \sin i$ --  the semi-major axis of the eclipsing
pair around the common center of gravity in astronomical units
determined using the LITE; $\epsilon_{BV}$ -- the difference
between observed TYCHO-2 $B-V$ color index and the SPBE (see the
text); $\log X$ = $\log f_x/f_{bol}$ -- the flux ratio of the
X-ray flux from the RASS survey to the bolometric flux (see
Section 4.3); Moon -- ``M'' signifies that the system can be
occulted by the Moon (within 5$^\circ$9' from the ecliptic); FLAGS
are explained in Table~\ref{flags}; Combined weight -- The sum of
weights of individual detection techniques (any detection counts
with weight 1.0, any possibility or suspicion with 0.5); Triple --
is our final evaluation of triplicity with ``Y'' = Yes, if sum of
weights is equal or larger than 1.0. \\
Systems without GCVS
designations: V2 Mon = TYC 4824-00153-1, V3 Mon = TYC
4835-01947-1,
V2 Peg = TYC 1720-00658-1.\\
}
\end{deluxetable}

%-----------------------------------------------------------------------------
\begin{deluxetable}{lp{14cm}}

%\small
\tabletypesize{\scriptsize}

\tabcolsep 3pt
%\tablecolumns{2}

\tablewidth{0pt}
%\tablewidth{320pt}
\tablenum{2}

\tablecaption{Explanation of FLAGS for different types of observations.
 Flags applicable to particular type of technique are given in
              parentheses \label{flags}}
\tablehead{
\colhead{FLAG}            & %1
\colhead{Definition}        \\ %2
}
\startdata
     &  Generic names: {\tt D} detection, {\tt S} strong suspicion of detection,
        {\tt P} possibility of a companion, {\tt N} no detection,
        {\tt -} not observed or insufficient data.\\
(1)  &  Visual binary present in the WDS Catalog ({\tt DPN}). "D"
        for systems with projected separation less than 2000 AU, "P" for systems with
        separation between 2000 and 20000 AU, "N" systems not present in WDS or with
        separation larger than 20000 AU. \\
(2)  &  CFHT AO observations ({\tt DPN-}). The flags same as in WDS.
        "-" here means that the system was not observed.\\
(3)  &  H59 ({\tt CGX-}) and H61 ({\tt ABCS-}) Hipparcos Catalog
        flags.
        ``D'' if X or G in H59, ``S'' if S or C in H61.\\
(4)  &  $\sigma_\mu$ ({\tt SN-}).
        ``S'' if proper-motion error higher than 3-$\sigma$ from the median. \\
(5)  &  Spectroscopic detection ({\tt DSN-}. Detection in broadening functions
        or through spectral features ({\tt DN-}). Large
        changes of systemic velocity were coded by ``S''.\\
(6)  &  LITE indication ({\tt DSN-}).
        {\tt D}: a stable orbit with more than 1.5 cycles covered,
        {\tt S}: a somewhat unstable solution or less than 1.5 cycles covered,
        {\tt -}: less than 25 {\it pe} or CCD minima or the
        interval of observations $<10$ years. \\
(7)  &  SPBE indication ({\tt SN-}).
        {\tt S} if the system is bluer than the corresponding blue envelope color. \\
(8)  &  RASS X-ray to visual flux ratio ({\tt LEN}).
        {\tt L} when a late-type companion is indicated by a large $f_X/f_{bol}$ ratio in
        early-type contact binary, {\tt E} when the ratio is below the
        lower envelope of detection limit indicating an
        early-type third component, otherwise {\tt N}. \\
\enddata
\end{deluxetable}

%-----------------------------------------------------------------------------
%Table with comments for individual systems

\begin{deluxetable}{lp{14cm}}

%\small
\tabletypesize{\scriptsize}

\tabcolsep 3pt
%\tablecolumns{3}

\tablewidth{0pt}
%\tablewidth{320pt}
\tablenum{3}

\tablecaption{Notes on selected contact binaries \label{notes}}

\tablehead{
\colhead{Name}            & %1
\colhead{Comments}        \\ %2
}
\startdata
AB And   & The third component is clearly seen in the LITE, but has
not been detected during the CFHT AO observations.\\
GZ And   & A multiple (trapezoidal) system, with colors and
magnitudes indicating that all visual components may be physically
associated. It does not have the Hipparcos parallax.\\
V410 Aur & The visual component, identical with that of
the spectroscopic detection, shows RV variability
indicating that it is possibly SB1.\\
44 Boo   & The closest contact binary.
The 206 year period of the visual orbit is reflected
in the LITE, but cannot be determined solely from this effect, so it is
considered as {\it not detected\/} by the LITE.\\
CK Boo   & The third component, detected independently
by the LITE and AO CFHT observations, is very probably the same body.
\citet{krez1991} detected a small, but definite ultraviolet
excess, so it is possible that the third component is a white dwarf.\\
AO Cam   & A very preliminary LITE solution has been given
by \citet{qian2005}. Spectral signatures of a
companion detected by \citet{dangelo2005}. \\
AH Cnc   & A member of M67.
A physical link within WDS~08517+1151 is uncertain.\\
VW Cep   & The third component has been detected by several techniques.
 The visual and LITE orbits still show an additional wobble.\\
EM Cep   & A member of the open cluster NGC~7160. The projected
separation of the components ($\rho/\pi$) is almost 50,000 AU,
which is much larger than for all remaining wide systems of this
sample. Hence, it is not considered as a genuine triple system.\\
EE Cet   & While the southern component of WDS~02499+0856 is a contact binary,
 unpublished DDO spectroscopic observations show that the northern component
 is a double-lines (SB2) binary hence it is a quadruple system.\\
BV Dra \& BW Dra & A well known quadruple system composed of two contact
  binaries. BW~Dra shows additional astrometric complications.\\
V345 Gem & The contact binary is the fainter component of a visual double.\\
AK Her   & The fainter visual component cannot be identified with the
component indicated by the LITE, so the system is possibly a quadruple one.\\
SW Lac   & Lines of the third component were detected spectrally
by \citep{HenryMki1998}; this is probably the same component
as seen in the CFHT AO data. The LITE indicates two companions, but
the visual companion does not seem to correspond to
any of the possible bodies suggested by the LITE solution. \\
XY Leo   & A member of a quadruple system detected by \citep{bard1987}.
  The companion binary is a BY~Dra-type binary system.
  The LITE orbit is exceptionally well defined.\\
AM Leo   & A member of relatively wide visual binary. A marginally
acceptable LITE solution for the
 contact binary was given by \citet{qian2005}.\\
VW LMi   & A quadruple system according to DDO spectral observations
 (to be published). The companion binary has an orbital period
 of 7.93 days while both binaries appear to revolve about the
 common center of gravity in 356 days. \\
V752 Mon & The contact binary is the fainter component
of the visual double. The brighter primary is a rapidly rotating A-type star
  obliterating signatures of the contact pair in the spectra. \\
V502 Oph & Lines of a third component were detected by \citep{HenryMki1998}.\\
V2388 Oph& A visual binary with the orbital period of only 8.9 years.\\
BB Peg   & \citet{dangelo2005} detected M dwarf signatures in the spectrum.
      An LITE analysis indicates a compatible body with a
      minimum mass of 0.18 M$_\odot$. \\
AW UMa   & A proper motion pair with BD+30~2164. The projected
     separation of the components is $d = \rho/\pi$ = 4,400 AU.
     The magnitudes and colors indicate a physical link.
The LITE and systemic velocities show
additional perturbations which may be related to another, closer companion. \\
TV UMi   & A quadruple system according to the DDO spectral
 observations (to be published).
 The second binary is a detached system with
 a highly eccentric orbit and a period of a few weeks.\\
AG Vir & The faint companion has been detected in the DDO spectra
(to be published).\\
AH Vir   & The visual companion cannot be identified with the
       component indicated by the LITE.\\
HT Vir   & Components of WDS~13461+0507 revolving in 274 years orbit are both
       binaries. While the component B is contact binary, the
       component A is a single-lined (SB1)
       binary with $P$ = 32.45 days \citep{ddo04}.
\enddata
\end{deluxetable}

%-----------------------------------------------------------------------------

\begin{deluxetable}{lccrrrrcc}

\tabletypesize{\scriptsize}

%\tablecolumns{8}
\tabcolsep 3pt
\tablewidth{0pt}
\tablenum{4}

\tablecaption{CFHT Adaptive Optics observations (1998). \label{cfht}}
\tablehead{
\colhead{Name} & \colhead{Comp.} & \colhead{$\rho$} & \colhead{$\theta$} & \colhead{$\Delta K$} &
\colhead{$\Delta H$} & \colhead{HJD} & FLAG & \colhead{Comments} \\
 & &  \colhead{[arcsec]} &  \colhead{[deg]} &  &  &  \colhead{--2450800} & & \\
 }
\startdata
AB   And       &    &  18.9&     306&   7.3&      &  219.04 &  P & uncertain  \\
GZ And\tablenotemark{a} & B  &   8.6&      36&  -0.2&  -0.3 & 219.10 & D & poor seeing \\
GZ   And       & C  &  13.3&     196&  -0.3&   0.0&  219.10 &   & poor seeing \\
\underline{GZ   And} & E  &  2.13&      33&   2.4&   2.6&  219.10 &   & poor seeing \\
\underline{GZ   And} & F  &  13.3&     134&   5.0&   5.5&  219.10 &   & poor seeing \\
\underline{CK Boo} &    &  0.12&     196&   2.8&  &  218.80 & D  &            \\
SX   Crv       &    &      &        &      &      &  218.75 & N  &            \\
V1073Cyg       &    &      &        &      &      &  218.98 & N  & poor seeing\\
YY   Eri       &    &      &        &      &      &   24.79 & N  &            \\
BV   Eri       &    &      &        &      &      &  219.12 & N  &            \\
AK   Her       &    &   4.5&     324&   2.0&      &  218.88 & D  & poor seeing\\
FG   Hya       &    &  16.5&     145&   4.0&      &   24.96 & P  &            \\
\underline{SW Lac} &    &  1.68&   85&   2.5& 2.6 &  219.02 & D  &            \\
UZ   Leo       &    &      &        &      &      &   25.05 & N  & poor obs   \\
XY   Leo       &    &      &        &      &      &   25.01 & N  &            \\
AP   Leo       &    &      &        &      &      &   25.07 & N  &            \\
UV   Lyn       &    &      &        &      &      &   24.99 & N  &            \\
V502 Oph       &    &      &        &      &      &  218.87 & N  & poor seeing\\
V566 Oph       &    &      &        &      &      &  218.89 & N  & poor seeing\\
V839 Oph       &    &   8.2&       8&   6.1&      &  218.92 & D  & poor seeing\\
V839 Oph       &    &  22.3&      36&   2.7&      &  218.92 &    & poor seeing\\
ER   Ori\tablenotemark{b} & C  &  0.17&     353&    2.2&      &   24.85 & D  & \\
ER   Ori                  & B  &  14.4&      10&    2.6&      &   24.85 &    & \\
\underline{U Peg}         &    &  4.05&     275&    5.2&      &  219.07 & D  & \\
AQ   Psc       &    &  21.5&      33&   4.2&      &   24.71 &  P &           \\
AQ   Psc       &    &  21.5&      33&   3.4&      &  219.08 &    &        \\
AQ   Psc       &    &  25.8&      65&   3.5&      &   24.72 &    &        \\
AQ   Psc       &    &  25.8&      65&   3.0&      &  219.08 &    &        \\
TY   Pup       &    &   8.4&       2&   3.8&      &   24.93 &  P &        \\
V781 Tau       &    &  30.0&     318&   3.0&      &   24.88 &  D &        \\
V781 Tau       &    &  14.6&     163&   8.0&      &   24.88 &    & uncertain  \\
AG   Vir       &    &      &        &      &      &   25.09 &  N &        \\
AH   Vir       &    &  1.71&      17&   1.6&      &   25.11 &  D &        \\
GR   Vir       &    &      &        &      &      &  218.82 &  N &        \\
\hline
%V417 Aql       &    &   8.5&      82&   4.8&      &  218.94 & poor seeing\\
%V417 Aql       &    &  12.5&     160&   5.0&      &  218.94 & poor seeing\\
%EL   Aqr       &    &      &        &      &      &  219.06 &            \\
\underline{AH Aur} &    &  3.18&      63&   4.2&   4.0&   24.90 & D &          \\
AH   Aur       &    &  17.2&      80&   7.1&   7.1&   24.90 &    & uncertain   \\
AH   Aur       &    &  21.8&     326&   6.6&   6.6&   24.90 &    & uncertain   \\
%AC   Boo      &    &      &        &      &      &  218.83 &    &             \\
TZ   Boo       &    &  10.6&     120&   6.9&      &  218.84 & D  & uncertain   \\
%VW   Boo      &    &  16.3&     259&   2.2&   2.4&  218.78 &    &             \\
%XY   Boo      &    &  17.1&      34&   7.3&      &  218.77 &    & uncertain   \\
%TT   Cet      &    &   7.8&      55&   4.3&   4.4&  219.09 &    &             \\
RW   Com       &    &      &        &      &      &   25.12 & N  & poor obs    \\
%RZ   Com      &    &      &        &      &      &   25.14 &    &             \\
DK   Cyg       &    &      &        &      &      &  218.99 & N  & poor seeing \\
V401 Cyg       &    &  18.0&     359&   1.2&      &  218.93 & P  & poor seeing \\
%SV   Equ      &    &      &        &      &      &  218.97 &    & poor seeing \\
UX   Eri       &    &      &        &      &      &   24.76 & N  & poor obs    \\
XZ   Leo       &    &  18.3&     112&   7.0&      &   25.03 & P  & uncertain   \\
VZ   Lib       &    &      &        &      &      &  218.86 & N  & poor seeing \\
\underline{V508 Oph} &    &  2.40&  19&  3.9&  4.2&  218.90 & D & poor seeing  \\
V508 Oph       &    &  13.9&     192&   4.6&   4.8&  218.90 &   & poor seeing  \\
BB   Peg       &    &      &        &      &      &  219.01 & N &              \\
%VZ   Psc      &    &      &        &      &      &  219.05 &   &              \\
\underline{RZ Tau} &    &  0.80&      43&   3.7&      &   24.81 & D &      \\
\underline{RZ Tau} &    &  0.80&      44&   3.9&      &  219.13 &   &      \\
\enddata
\tablecomments{
The second part of the table below the horizontal line contains binaries
fainter than $V=10$.\\[1ex]
New detections at separations $<5$ arcsec are underlined.\\[1ex]
Comments on poor observations:\\
Poor seeing: The adaptive optics system
worked poorly when the natural seeing
was worse than 1 arcsec;\\
Poor observation: The central star (contact binary)
saturated in some images;\\
Uncertain: The detected companion was very faint, at the limit of detection.\\[1ex]
$\Delta K$ and $\Delta H$ magnitude differences relative
to the central (contact binary) star were measured
with accuracy of about 0.1 -- 0.2 mag.\\
Column FLAG gives our rating of the target corresponding to that in Table~\ref{summary}.
The flag is given only in the line of first appearance of target.\\
}
\tablenotetext{a}{GZ And: Components B and C, as identified in
\citet{walk1973}. The component D was outside the field of view.
Components E and F are new.}
\tablenotetext{b}{ER Ori: The more distant companion known
before is called B while the physical companion discovered
by \citet{goe1994} and confirmed here is called C.}
\end{deluxetable}

%----------------------------------------------------------------------------------

\begin{deluxetable}{lclclcccccrcclcc}

\tabletypesize{\scriptsize}

\tabcolsep 1.5pt

%\tablecolumns{17}

\tablewidth{0pt}
%\tablewidth{320pt}
\tablenum{5}

\tablecaption{LITE orbits, estimated masses and angular distances of multiple
              components. \label{lite}}
\tablehead{
\colhead{Name}            & %1
\colhead{$P_3$}           & %2
\colhead{$e_3$}           & %3
\colhead{$a_{12} \sin i$} & %4
\colhead{$f(m)$}          & %5
\colhead{Cycle}           & %6
\colhead{$M_{12} \sin^3 i$} &%7
\colhead{$i$}             & %8
\colhead{$M_{12}$}        & %9
\colhead{$M_3$}           & %10
\colhead{$\pi$}           & %11
\colhead{$\alpha$}        & %12
\colhead{$\Delta V_0$}    & %13
\colhead{$A$}             &
\colhead{FLAG}            &
\colhead{Ref.}            \\%14
                          & %1
\colhead{[years]}         & %2
                          & %3
\colhead{[AU]}            & %4
\colhead{[M$_\odot$]}     & %5
                          & %6
\colhead{[M$_\odot$]}     & %7
\colhead{[deg]}           & %8
\colhead{[M$_\odot$]}     & %9
\colhead{$i=90^\circ$}    & %10
 \colhead{[mas]}          & %11
 \colhead{[mas]}          & %12
 \colhead{[km.s$^{-1}$]}  & %13
 \colhead{[days]}         &
                          &
                          \\%14
}
\startdata
AB And  & 59.1(10) & 0.12(3) & 3.77(6)& 0.0153(9)  & 1.72 & 1.648 & 86.8 & 1.656 & 0.40 &  8.34 & 160 & 2.09 & 0.0433 & D & (1) \\
BX And  & 62(9)    & 0.3(3)  & 2.2(3) & 0.0027(11) & 1.68 &       & 74.5 &       &      &  6.60 &     & 2.20 & 0.0254 & S & (2) \\
CK Boo  & 21.3$^f$ & 0.96(7):& 5(4)   & 0.3(6)     & 1.36 & 1.171 & 56.6 & 2.012 & 1.50 &  6.38 & 300:&  40: & 0.0244 & S &     \\
VW Cep  & 27.8(2)  & 0.54(4) & 2.28(6)& 0.0153(12) & 2.80 & 1.280 & 63.6 & 1.781 & 0.42 & 36.16 & 390 & 5.82 & 0.0246 & D & (4) \\
AK Her  & 57.8(8)  & 0.16(6) & 1.65(6)& 0.00134(16)& 1.81 &       & 80.8 &       &      & 10.47 &     & 1.72 & 0.0190 & D & (1) \\
SW Lac3 & 93.1$f^3$& 0.71(1) &15.2(2) & 0.405(9)   & 1.27 & 2.101 & 80.2 & 2.196 & 1.89 & 12.30 & 400 &13.82 & 0.1294 & D & (6) \\
SW Lac4 & 28.2(1)  & 0.51(5) & 2.58(9)& 0.021(3)   & 4.20 & 2.101 & 80.2 & 2.196 & 0.54 & 12.30 & 160 & 6.33 & 0.0278 & D & (6) \\
XY Leo  & 19.59(5) & 0.09(3) & 4.08(4)& 0.1774(58) & 3.01 & 1.046 & 65.8 & 1.378 & 1.00 & 15.86 & 140 &12.36 & 0.0470 & D & (7) \\
V566 Oph& 23.1(5)  & 0.6(2)  & 1.2(2) & 0.0031(17) & 2.60 & 1.743 & 79.8 & 1.828 & 0.24 & 13.98 & 140 & 4.01 & 0.0111 & D &     \\
ER Ori  & 52(3)    & 0.77(5) & 5.7(5) & 0.068(18)  & 1.40 & 2.503 & 87.5 & 2.510 & 0.93 & -6.68 &     &12.69 & 0.0418 & S & (8) \\
AW UMa  & 16.8(5)  & 0.46(22)& 0.44(6)& 0.00030(12)& 2.33 & 1.815 & 78.3 & 1.933 & 0.11 & 15.13 & 120 & 1.77 & 0.0051 & D &(10) \\
AH Vir  & 41.1(15) & 0.89(5) & 4.2(8) & 0.042(24)  & 1.88 & 1.889 & 86.5 & 1.900 & 0.65 & 10.86 & 180 &13.21 & 0.0219 & D &(11) \\
\hline
TY Boo  & 55.8(27) & 0.26(7) & 4.1(3) & 0.0214(47) & 1.38 & 1.637 & 77.5 & 1.759 & 0.47 &       &     & 4.48 & 0.0457 & S &     \\
TZ Boo  & 34.0(5)  & 0.58(4) & 6.3(2) & 0.217(23)  & 2.30 & 0.691 &      &       &      &  6.76 &     &13.50 & 0.0708 & D &     \\
V523 Cas& 86(5)    & 0.08(7) & 4.8(4) & 0.015(4)   & 1.17 & 1.110 & 83.8 & 1.130 & 0.31 &       &     & 3.35 & 0.0558 & S & (3) \\
GW Cep  & 13.5(5)  & 0.58(28)& 0.69(8)& 0.0018(7)  & 3.06 &       & 84.0 &       &      &       &     & 3.72 & 0.0079 & D &     \\
UX Eri  & 42.8(13) & 0.6$f^3$& 3.9(3) & 0.033(7)   & 1.74 & 1.798 & 79.0 & 1.901 & 0.59 &  6.57 & 110 & 6.80 & 0.0366 & S &     \\
V829 Her& 10.9(6)  & 0.26(32)& 0.9(2) & 0.006(4)   & 1.37 & 1.062 & 57.0 & 1.800 & 0.30 &       &     & 5.16 & 0.0102 & S & (5) \\
BB Peg  & 20.4(4)  & 0.32(21)& 0.83(9)& 0.0014(5)  & 3.59 & 1.862 & 79.9 & 1.951 & 0.19 &  3.02 &  30:& 2.59 & 0.0092 & S &     \\
EQ Tau  & 50.2(18) & 0.37(12)& 3.9(4) & 0.024(7)   & 1.22 & 1.749 & 86.6 & 1.752 & 0.49 &       &     & 5.21 & 0.0390 & S & (9) \\
\enddata
\tablecomments{
The second part of the table below the horizontal line contains binaries
fainter than $V=10$.\\[1ex]
System: GCVS name;
$P_3$: The orbital period of the 3rd body in years;
$e_3$: The orbital eccentricity; $a_{12} \sin i$: The
semi-major axis of the eclipsing binary orbit around
the common center of gravity; $f(m)$: The mass function;
Cycle: The number of cycles of the third body observed;
$M_{12} \sin^3 i$: The minimum mass of the contact pair; $i$:
the orbital inclination of the contact pair; $M_{12}$: The mass of the contact pair;
$M_3$: The minimum mass of the third component;
$\pi$: The parallax of the system; $\alpha$: The estimated angular
separation of the third component; $\Delta V_0$: The mean difference
in radial velocities, $A$: The full amplitude of the systemic velocity and of the
(O--C) changes, respectively, as predicted
by LITE. The superscript ``f'' on $P_3$ or $e_3$ means that the parameter
was fixed at an optimum value.
For SW Lac we give the separate solution for third and fourth component.\\
References (given only for stars previously known or suspected to display
LITE or cyclic period variations):\\(1) -- \citet{borhe1996},\\ (2) -- \citet{demir1993},\\
(3) -- \citet{samec2004},\\ (4) -- \citet{prib2000},\\
(5) -- \citet{erdem2004},\\ (6) -- \citet{prib1999a},\\
(7) -- \citet{yakut2003},\\ (8) -- \citet{kim2003},\\
(9) -- \citet{privan2002},\\ (10) -- \citet{prib1999b},\\ (11) -- \citet{hobapen1999}
}
\end{deluxetable}

\end{document}